\documentclass[aps,superscriptaddress,prd,
onecolumn,
floatfix,
nofootinbib,
amsmath,amssymb,amsfonts]{revtex4-2}

\usepackage{mathrsfs}
\usepackage[pdftex]{graphicx}
\usepackage{color,float}
\usepackage[colorlinks,linkcolor=blue,anchorcolor=violet,citecolor=red]{hyperref}
\usepackage{physics}
\usepackage{comment}
\usepackage{bm}

\newcommand{\pa}{\partial}
\newcommand{\veps}{\varepsilon}

\newcommand{\bs}[1]{\boldsymbol{#1}}

\begin{document}
\title{Particle Creation and Entanglement in Dispersive  Model with Step Velocity Profile}

\author{Yuki Osawa}
\email{osawa.yuki.e8@s.mail.nagoya-u.ac.jp}
\author{Yasusada Nambu}
\email{nambu@gravity.phys.nagoya-u.ac.jp}
\affiliation{Department of Physics, Graduate School of Science, Nagoya
University, Chikusa, Nagoya 464-8602, Japan}

\date{April 19, 2022} 
\begin{abstract}
  We investigate particle creation and entanglement structure in
    a dispersive model with subluminal dispersion relation. Assuming the step
    function spatial velocity profile of the background flow, mode
    functions for a massless scalar field is exactly obtained by the
    matching method. Power spectrums of created particles are calculated
    for the subsonic and the transsonic flow cases. For the
    transsonic case, the sonic horizon exists and created particles
    show the Planckian distribution for low frequency region but the
    thermal property disappears for high frequency region near the
    cutoff frequency introduced by the non-linear dispersion. For the
    subsonic case, although the sonic horizon does not exist, the
    effective group velocity horizon appears due to the non-linear
    dispersion for high frequency region and approximate thermal
    property of the power spectrum arises. Relation between particle
    creation and entanglement between each  mode is also discussed. 
\end{abstract}

\maketitle

\section{Introduction}
Quantum field theory in black hole spacetimes predicts emission of
thermal Hawking radiation from black holes, of which temperature is
given by the surface gravity at their event horizons
\cite{Hawking1,Hawking2} . This property of Hawking radiation from
black holes suggests that black holes behave as a kind of
thermodynamical objects and the theory of black hole thermodynamics is
formulated \cite{Bekenstein}.  Thermal property of Hawking radiation
leads to a problem called the information loss paradox, and for deeper
understanding of this issue, analysis of
entanglement between Hawking mode (Hawking radiation) and its partner
mode has been done to investigate the quantum informational aspect of the black
hole evaporation \cite{Page}. Investigations of  entanglement for
analog models of black holes also have been done recently
\cite{Busch,Bruschi,Isoard,nambu}.

The original Hawking's scenario implies that low energy radiations are
originated from high energy region above the Planckian scale, at which
quantum gravitational physics will become important. Thus it is
crucial to clarify effect of the Planckian scale cutoff on the thermal
property of Hawking radiation (the trans-Planckian problem)
\cite{Jacobson1,BMPS}.  To resolve this problem, it is necessary to
consider the origin of the particle radiated from the black hole.  If
we consider time reversed evolution of emission of Hawking radiation,
frequencies of emitted radiations increases exponentially as they
approach the black hole horizon and exceeds the Planckian frequency,
beyond that frequency, quantum effect of gravity may become important.  To investigate such a
situation, Unruh proposed sonic analog of black holes
\cite{Unruh1,Unruh2}; he found that the equation of sonic waves in
moving fluid has the same form as a massless scalar field in curved
spacetimes, of which metric has the similar structure as black hole
spacetimes. The acoustic metric corresponds to the black hole
spacetime with Painlev\'e coordinates
\begin{equation}
  ds^2=-c^2dt^2+(dx-v(x)dt)^2,\quad v(x)=-c\sqrt{x_s/x}
\end{equation}
where $x_s$  is the  Schwarzschild radius
 and $c$ is the light velocity.  The cutoff wave number $k_0$
is introduced as the distance between atoms constituting fluid.  He
numerically calculated the power spectrum of the radiation for the
analog black hole with the high frequency cutoff and has shown that
this cutoff does not affect the spectrum of Hawking radiation in a low
frequency region.

Owing to introduction of the frequency cutoff, the Lorentz invariance
of the system is broken and additional wave modes associated with the
cutoff appear.  Owing to these modes, Hawking radiation in analog
black holes is emitted by a process that the Planckian modes
transformed into the low energy modes (mode conversion).  If the
velocity profile $v(x)$ is a slowly changing function of the spatial
coordinate, a lot of analyses have been done so far based on the WKB
method. These studies show that for $\omega, \kappa \ll k_0$
where $\kappa$ is the first derivative of the velocity profile at the
sonic horizon, the temperature of the radiation is given by
$\kappa/(2\pi)$ and it coincides with Hawking's
results~\cite{BMPS,UnruhSchutzhold,LeonhardtRobertson,Robertson,CorleyJacobson}.

If the velocity profile is not slowly changing function of the spatial
coordinate and expected temperature of analog black holes is high,
several studies of particle creations in analog models
\cite{Corley,Finazzi,MFR,Robertson,Coutant} show that the spectrum of the
radiation is determined not only by the first derivative of the fluid
velocity at the sonic horizon, but it also depends on other parameters
including the frequency cutoff.  Although mechanism of radiation from
analog black holes with high temperature does differs from the
original Hawking radiation, it is important to study entanglement of
involved modes for such a case to understand effect of the frequency
cutoff on emission mechanism of radiation. In this paper, we consider
an analog model with a step function velocity profile of the
background flow, and apply the step discontinuous method introduced by
\cite{MFR,Finazzi,Robertson} to evaluate Bogoliubov coefficients. Then we
calculate number density of created particles and entanglement
negativity between involved modes.  The present study may have
overlaps with the analysis by X. Busch and R. Parentani \cite{Busch},
in which entanglement structure for high temperature analog black hole
was studied.  Our analysis differs from their work in several points;
first, we calculated the multipartite entanglement between modes.
Second, we consider two different types of analog spacetimes; the one
with a sonic horizon, and the other without a sonic horizon.

The paper is organized as follows. In Section II, we shortly review
particle creation in analog system with the dispersive media. In
Section III, we determine the Bogoliubov coefficients for the step function
velocity profile.  In Section IV, we investigate entanglement
structure of the in-vacuum state.  In Section V, we show results of
our numerical calculation.  Section VI is devoted to summary and
conclusion.  We used the unit $c=\hbar=G=1$ throughout this paper.

\section{Wave modes for steep velocity profile}
We consider wave modes of an analog model with dispersive media. We
adopt the following wave equation of a massless real scalar field 
\begin{equation}
        (\pa_{t}+\pa_x v(x))(\pa_{t}+v(x)\pa_x)\phi(x,t)
        =c_{s}^2(-i \pa_x)\,\pa_x^2\phi(x,t).
        \label{eq:weq}
\end{equation}
This is the equation for sonic waves in a moving fluid  with a
position dependent velocity profile $v(x)$ and the sound velocity
$c_{s}(k)$ with the wave number $k=-i\pa_x$. In the following, we
assume that the velocity of the fluid has the step function
profile\footnote{The step function is defined by
\begin{equation*}
  \theta(x)=
  \begin{cases}      0 & (x<0)\\
                        1/2 & (x=0)\\
                        1 &(x>0).
  \end{cases}
\end{equation*}
} 
\begin{equation}
v(x)=V_-+(V_+-V_-)\, \theta(x), \quad V_{\pm}<0, 
\label{eq:v}
\end{equation}
and the flow velocity in $x>0$ region is
subsonic $V_+>-1$. We assume subluminal dispersion $c_{s}^2(k)=1-k^2/k_0^2$ with the cutoff
of wave number $k_0$. The sonic horizon exists at $x=0$ if
$v(0)=(V_++V_-)/2<-1$ and in such a case, the region $x<0$ becomes
supersonic (inside of the sonic horizon). We will see in the next
subsection A, even when whole region is subsonic and there is no
sonic horizon, owing to dispersive property of the fluid for high
frequency, there exists an effective horizon (group velocity horizon)
which has the similar property as the sonic horizon.

\subsection{Mode Functions}
\begin{figure}[H]
  \centering
  \includegraphics[width=0.35\linewidth]{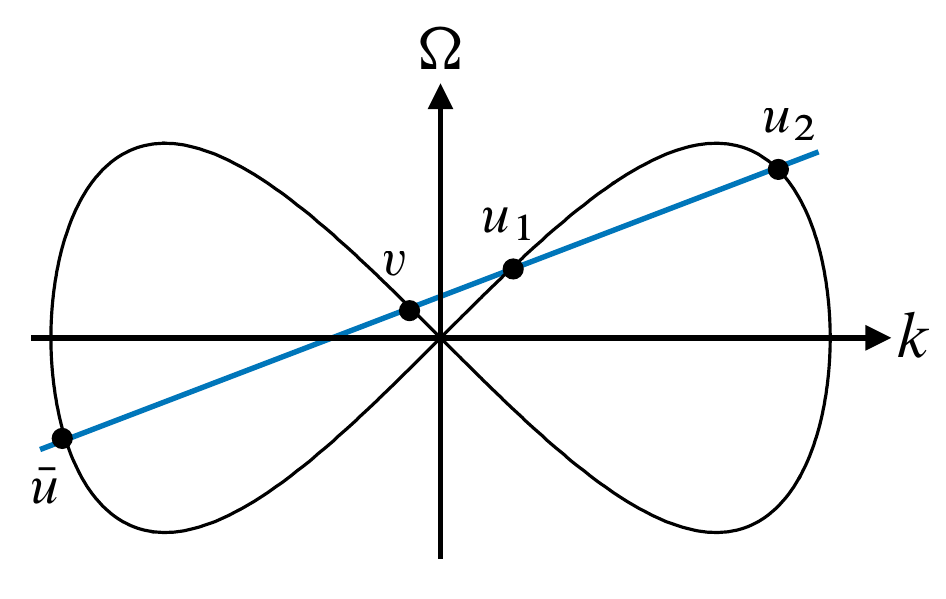}
  \caption{Dispersion diagram with the subluminal dispersion for the
    subsonic case.  A straight line represents
    $\Omega=\omega-V_{\pm}\,k$. The dispersion relation has four roots
    which define modes. We name them as $\bar u, v, u_1, u_2$ modes in
    the increasing order of $k$.}
      \label{fig:disp}
\end{figure}

\begin{figure}[H]
  \centering
  \includegraphics[width=0.8\linewidth]{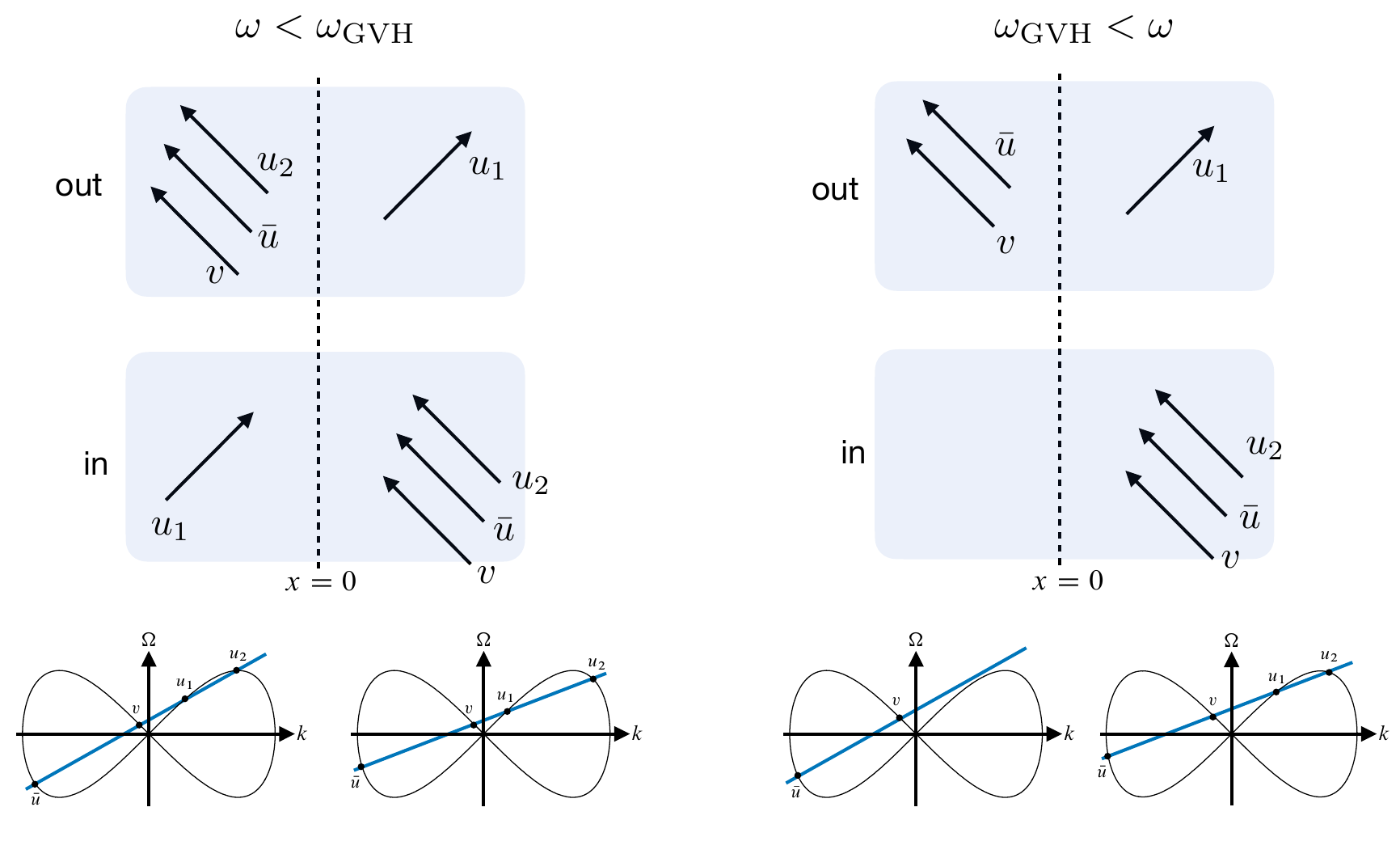}
  \caption{Dispersion diagrams and modes for the subsonic case. For
    $\omega<\omega_\text{GVH}$, the in-state and the out-state contain
    independent four modes. For $\omega_\text{GVH}<\omega$, there in
    no right moving modes in $x<0$ and the in-state and the out-state
    contain independent three modes.}
      \label{fig:sub-diagram}
\end{figure}

\begin{figure}[H]
  \centering
  \includegraphics[width=0.4\linewidth]{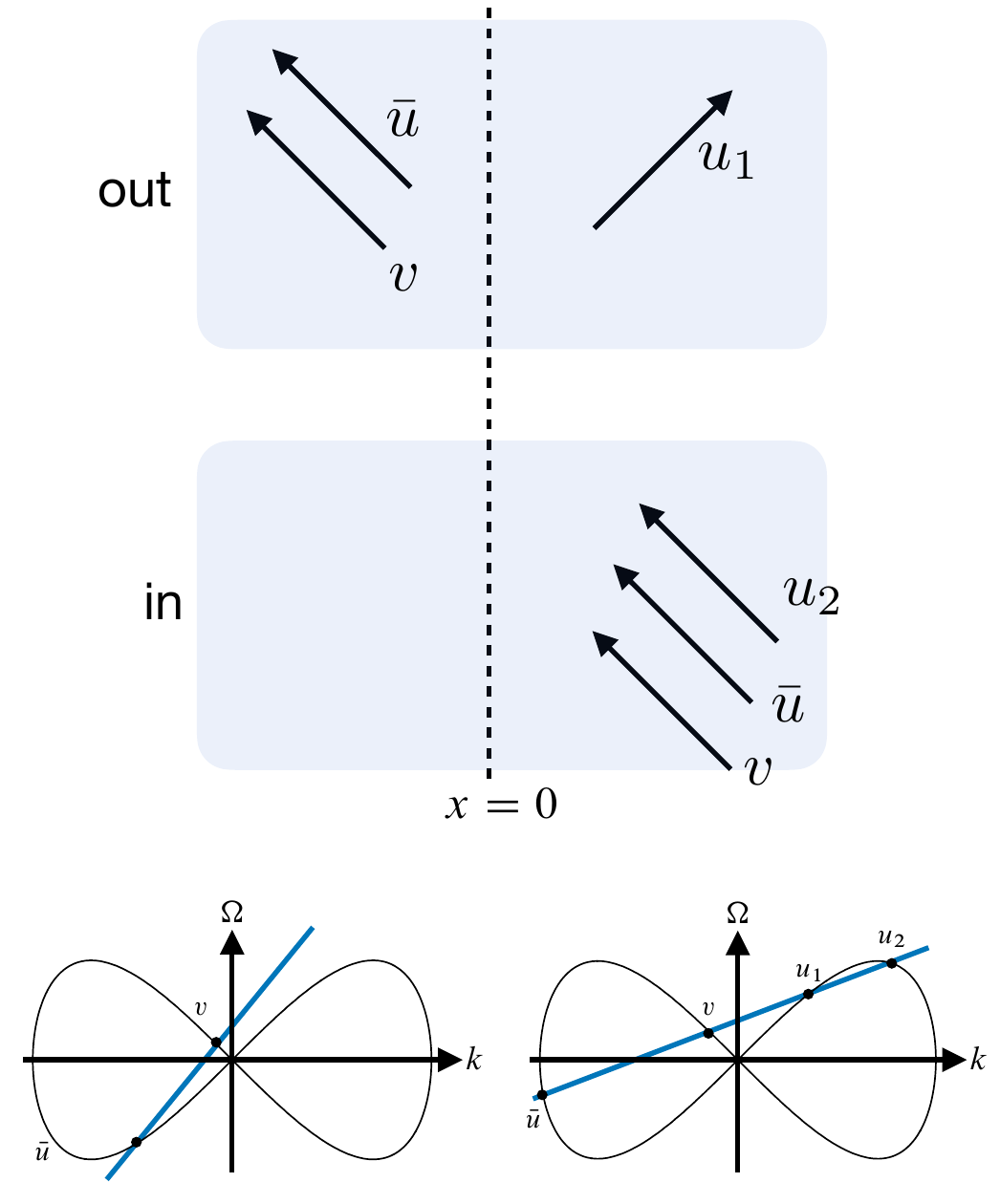}
  \caption{Dispersion diagrams and modes for the transsonic
    case. There is no right moving modes in $x<0$.}
      \label{fig:trans-diagram}
\end{figure}

For the wave equation Eq.~\eqref{eq:weq} with the subluminal
dispersion and the velocity profile Eq.~\eqref{eq:v}, by assuming
$\phi\propto e^{-i\omega t+ikx}~(x\neq 0)$, the dispersion relation is
obtained as
\begin{equation}
\omega-k V_\pm=\pm |k|\sqrt{1-\left(\frac{k}{k_0}\right)^2},\quad \omega>0.\quad\text{(signs in no particular order)}
\label{eq:dispersion}
\end{equation}
We denote $k^{\pm}_{i}(\omega)$ as solutions of the dispersion
relation corresponding to $V_{\pm}$, where the index $i$ denotes a
label to distinguish modes.  Since the spacetime is static and the
wave equation does not contain explicit time dependence, a plane wave
with a frequency $\omega$ does not couple to other plane waves with
different $\omega$. We call $\omega$ as the laboratory frequency and
$\Omega:=\omega-k V_\pm$ as the comoving frequency. Note that the
comoving frequency is not conserved and changes its value depending on
$x$.  Solutions of the dispersion relation and corresponding modes in
our setup are shown in Fig.~\ref{fig:disp} using a dispersion diagram;
the vertical axis is the comoving frequency and the horizontal axis is
the wave number. The curve in the diagram represents the right hand
side of the dispersion relation $\pm |k|\sqrt{1-(k/k_0)^2}$, and the
straight line represents the left hand side of the dispersion relation
$\omega-k V_\pm$.
\noindent
From this diagram, we can identify modes as four real roots
$k_{\bar{u}}$, $k_{v}$, $k_{u_1}, k_{u_2}$ in the increasing
order of $k$. For larger values of $\omega$ or $|V_{\pm}|$,
we will have only two real solutions
$k_{\bar{u}}, k_{v}$ with $k<0$ and two pure imaginary
solutions. One of the imaginary solutions corresponds to the decaying
mode and the other imaginary solution corresponds to the growing
mode. For the velocity profile with the step function \eqref{eq:v},
  mode functions $\phi^\pm_i$ are plane waves
\begin{equation}
  \phi^\pm_{\omega,i}(t,x)=e^{-i\omega t}\,C_{\omega,i}^\pm
  \exp\left(i k^\pm_i(\omega)x\right),\quad i=\bar u, v, u_1, u_2,
\end{equation}
where $C_{\omega,i}^\pm$ are normalization constants.  We call 
$u_{2}^{+}, \bar u$ as  Planckian modes and $u_{1}^{+}, v$ as  non-Planckian
modes.  The Planckian modes appears due to non-linearity of the
dispersion relation. On the other hand, the non-Planckian modes exist
even for linear dispersion without the cutoff effect. The naming of
modes depends on values of $\omega$ and $V_\pm$. With the increase of $\omega$,
the non-Planckian mode $u_1$ approaches the Planckian $u_2$ mode. In
such a situation, we call $u_1$ as the sub-Planckian mode.  The group
velocity of each mode is given by
\begin{equation}
v_g=\left(\frac{d k(\omega)}{d\omega}\right)^{-1}.
\end{equation} 
We present behavior of modes for the subsonic case
(Fig.~\ref{fig:sub-diagram}) and the transsonic case
(Fig.~\ref{fig:trans-diagram}). The in-modes are defined as modes with
negative group velocity for $x>0$ and positive group velocity for
$x<0$ (incoming to $x=0$ from $x=\pm\infty$). The out-modes are
defined as modes with positive group velocity for $x>0$ and negative
group velocity for $x<0$ (outgoing from $x=0$ towards $x=\pm\infty$).

For the subsonic case (Fig.~\ref{fig:sub-diagram}), we assume
$-1<V_-<V_+<0$. For sufficiently small $\omega$, there are four in and
out modes (left panel in Fig.~\ref{fig:sub-diagram}). If we increase
$\omega$, the mode $u_1$ and the mode $u_2$ in $x<0$ coalesce at the
critical frequency $\omega_\text{GVH}$, and above this frequency, we
have only two out-modes in $x<0$ (right panel in
Fig.~\ref{fig:sub-diagram}). For $\omega_\text{GVH}<\omega$, $x=0$
behaves as a sonic horizon because there exists no right moving modes
in $x<0$ and this region effectively becomes the supersonic region. We
call this effective horizon as the group velocity horizon
(GVH) \cite{Robertson}. There are three
in-modes in $x>0$ ($u_2,\bar u, v$), and one out-modes in $x>0$
($u_1$) and two out-modes in $x<0$ ($\bar u, v$).
For the transsonic case $V_-<-1<V_+<0$ (Fig.~\ref{fig:trans-diagram}), $x=0$ is the
sonic horizon. There are three in-modes in $x>0$ ($u_2, \bar u, v$),
and two out-modes in $x<0$ ($\bar u, v$) and one out-mode in $x>0$
($u_1$). There exists no right-moving modes in the supersonic region
$x<0$.

The incoming Planckian mode $u_2^+$ is reflected at the sonic horizon
or the GVH. And it is transformed to the outgoing sub-Planckian mode
$u_1^+$. This process is called the mode conversion \cite{CorleyJacobson}. At the same time,
the Planckian mode $\bar{u}^+$ is transformed to the sub-Planckian
mode $\bar{u}^-$ for the transsonic case.  More detailed discussion
on modes for the slowly varying velocity profile can be found in
\cite{UnruhSchutzhold,LeonhardtRobertson,Robertson}.

\subsection{Quantization and vacuum state}

We quantize a classical field obeying the field equation
Eq.~(\ref{eq:weq}).  The action for the field is given by
\begin{align}
        S=\int dt\, dx \Bigl[\left|(\pa_t+v\,\pa_x)\phi\right|^2-\left|c_{s}(i\pa_x)\,\pa_x\phi\right|^2\Bigr],
\end{align}
and the conjugate momentum $\pi(t,x)$ for $\phi(t,x)$ is given by
\begin{align}
        \pi(t,x)=(\pa_t+v\,\pa_x)\phi(t,x).
\end{align}
The canonical commutation relation between quantized fields is imposed
as
\begin{align}
        \left[\hat{\phi}(t,x),\hat{\pi}(t,y)\right]=i\,\delta(x-y).
\end{align}
The Klein-Gordon inner product on $t=\text{const.}$ surface for
solutions $\phi_1,\phi_2$ of the field equation is defined by
\begin{align}
        (\phi_1,\phi_2):=-i\int dx\,\left(\phi_1D\phi_2^*-\phi_2^*D\phi_1\right)=-i\int_\Sigma dx~ \left(\phi_1\pi_2^*-\phi_2^*\pi_1\right),
\end{align}
where $D=\pa_{t}+v\,\pa_{x}$ and the inner product is conserved
\begin{align}
        \frac{d(\phi_1,\phi_2)}{dt}=0.
\end{align}
With the Klein-Gordon inner product, we can define creation and
annihilation operators associated with  the positive norm solution
$\{\phi_i\}$ of the wave equation by
\begin{align}
        \hat{a}(\phi_i)=(\phi_i,\hat{\phi}),\quad\hat{a}^\dag(\phi_i)=-(\phi_i^*,\hat{\phi}).
\end{align}
This set of creation and annihilation operators satisfies the following commutation relations:
\begin{equation}
        [\hat{a}(\phi_i),\hat{a}^\dag(\phi_j)]=(\phi_i,\phi_j),\quad
        [\hat{a}(\phi_i),\hat{a}(\phi_j)]=-(\phi_i,\phi_j^*),\quad
        [\hat{a}^\dag(\phi_i),\hat{a}^\dag(\phi_j)]=-(\phi_i^*,\phi_j).
\end{equation}
Thus if we choose a basis with ortho-normal condition, our creation
and annihilation operators satisfy the standard commutation
relation for the creation and annihilation operators.  In general, it
is not easy to construct exactly the ortho-normal basis with respect
to the Klein-Gordon inner product.  However, for the step function
velocity profile, as all modes are represented by plane waves, it is
easy to identify positive frequency modes which define a vacuum state.
We note that the mode functions for $\bar{u}^{\pm}$ have negative norms
and other modes have positive norms.  From a vacuum state,
multi-particle states are constructed by acting the creation operator
on the vacuum state.  We have two kinds of vacuum states.  The
in-vacuum state $|0_\text{in}\rangle$ is the state with no particle at
$t\rightarrow-\infty, x\rightarrow\pm \infty$
\begin{align}
        \hat{a}(\phi_i^\text{in})|0_\text{in}\rangle=0,
\end{align}
where $\phi_i^\text{in}$ is the positive frequency mode function of
the in-state for $i=u_{1}^{-},\bar{u}^+,v^+,u_2^+$ (sub-sonic case
with $\omega<\omega_\text{int}$), and $i=u_2^+, \bar u^+, v^+$
(sub-sonic case with $\omega_\text{int}<\omega$ or trans-sonic case).
The out-vacuum state $|0_\text{out}\rangle$ is the state with no
particle at $t\rightarrow+\infty, x\rightarrow\pm\infty$,
\begin{align}
        \hat{a}(\phi_i^\text{out})|0_\text{out}\rangle=0,
\end{align}
where $\phi_i^\text{out}$ is the positive frequency mode function of
the out-state for $i=\bar{u}^-,v^-, u_{2}^{-}, u_1^+$ (sub-sonic case
with $\omega<\omega_\text{int}$), and $i=\bar u^-, v^-, u_1^+$
(sub-sonic case with $\omega_\text{int}<\omega$ or trans-sonic case).
In general, these two vacuum state are not equal and the number of the
out-state particles in the in-state vacuum is
\begin{align}
       \langle0_\text{in}|\hat{a}^{\dag}(\phi_i^\text{out})\,\hat{a}(\phi_i^\text{out})|0_\text{in}\rangle\neq0.
\end{align}
This implies  particle creation occurs at $x=0$.

The filed operator is expanded as
\begin{equation}
\hat\phi(t,x)=\sum_i\left(\hat a(\phi_i^\text{in})\phi_i^\text{in}+\text{(h.c.)}\right)=\sum_i\left(\hat a(\phi_i^\text{out})\phi_i^\text{out}+\text{(h.c.)}\right),
\label{eq:fop}
\end{equation}
and creation and annihilation operators are represented as
\begin{equation}
 \hat a(\phi_i^\text{in,out})=(\phi_i^\text{in,out},\hat \phi),\quad \hat a^\dag(\phi_i^\text{in,out})=-(\phi_i^\text{in,out}{}^*,\hat \phi).
 \label{eq:ainout}
\end{equation}
\section{Bogoliubov coefficients}
We can  analytically determine a relation between the in-mode state and the out-mode state
for the wave equation with the step function velocity profile.

\subsection{Matching method}
By separating time dependence of the wave function as
$\propto e^{-i\omega t}$ in Eq.~\eqref{eq:weq}, the wave equation
becomes the following ordinary differential equation
\begin{align}
        \label{eq:weq2}
        (-i\omega+\pa_x v(x))(-i\omega+v(x)\,\pa_x)\phi(x)
        =\left(1+\frac{1}{k_0^2}\pa_{x}^{2}\right)\pa_{x}^{2}\,\phi(x)
\end{align}
with the velocity profile given by Eq.~\eqref{eq:v}. For $x\neq 0$,
the solution of this equation is superposition of plane waves
$\exp(i k_i^+ x)$ for $x>0$ and $\exp(i k_i^- x)$ for
$x<0$. Coefficients of superposition are determined by matching
conditions at $x=0$.
Let us denote $\phi_\pm$ as the solution of Eq.~\eqref{eq:weq2} for $x\gtrless0$.
We impose  matching conditions between $\phi_+$ and $\phi_-$ at $x=0$ as follows.  We require
continuity condition of $\phi$ at $x=0$ up to the second spatial
derivative to ensure the well-behaved wave function. Additional condition is obtained by integrating both
sides of the wave equation in the range $-\veps<x<\veps$, and taking
$\veps\to0$:
\begin{equation}
        \label{eq:1-8}
        -i\omega(V_+-V_-)\phi(0)+(V_+^2-V_-^2)\,\pa_x\phi(0)=\frac{1}{k_0^2}\left[\pa^3_x\,\phi^+(0)-\pa^3_x\,\phi^-(0)\right].
\end{equation}
After all, we require the following four matching conditions
\begin{align}
        &\phi^+(0)=\phi^-(0),\quad
        \pa_x\phi^+(0)=\pa_x\phi^-(0),\quad
        \pa^2_x\phi^+(0)=\pa^2_x\phi^-(0), \notag \\
        &\pa^3_x\phi^+(0)=\pa^3_x\phi^-(0)
        -{k_0^2}(V_{+}-V_{-})\left\{i\omega\phi(0)-(V_++V_-)\pa_x\phi(0)\right\}.
        \label{eq:matching}
\end{align}
Then the wave function 
$\phi(x)=\phi_+(x)\theta(x)+\phi_-(x)\theta(-x)$ is the global
solution of the wave equation \eqref{eq:weq2}.

\subsection{Bogoliubov coefficients}
By using the matching formula
Eq.~\eqref{eq:matching}, we can construct
$\phi^{+}_i(x)$ defined for $x>0$ connected to the plane
wave $e^{ik_{i}^{-}x}$ for $x<0$. 
$\phi^+_i(x)$ can be expressed as
\begin{equation}
  \label{eq:2-1}
  \phi^+_i(x)=\sum_{j=1}^4\alpha_{ij}\exp(ik^+_jx)
\end{equation}
with superposition coefficients $\{\alpha_{ij}\}$. Wave numbers
$\{k^+_j(\omega)\}$ are determined by Eq.~(\ref{eq:dispersion}).  The matching
formula Eq.~\eqref{eq:matching} yields the following
equations for $\{\alpha_{ij}\}$:
\begin{align}
         &\sum_j\alpha_{ij}=1,\quad
         \sum_j\alpha_{ij}\,k^+_j=k^-_{i},\quad
        \sum_j\alpha_{ij}\left(k^+_j\right)^2=\left(k^-_{i}\right)^2,\\
        &\sum_j\alpha_{ij}\left(k^+_j\right)^3=\left(k^-_{i}\right)^3+{k_0^2}\,(V_{+}-V_{-})\left\{\omega-(V_++V_-)\,k^-_{i}\right\}.
\end{align}
By solving these relation for $\alpha_{il}$, we obtain
\begin{align}
        \label{eq:2-6}
        \begin{pmatrix}\alpha_{i1}\\\alpha_{i2}\\\alpha_{i3}\\\alpha_{i4}\end{pmatrix}=
     \begin{pmatrix}
                        -{B_1}/{A_1}&{C_1}/{A_1}&-{D_1}/{A_1}&{1}/{A_1}\\
                        -{B_2}/{A_2}&{C_2}/{A_2}&-{D_2}/{A_2}&{1}/{A_2}\\
                        -{B_3}/{A_3}&{C_3}/{A_3}&-{D_3}/{A_3}&{1}/{A_3}\\
                        -{B_4}/{A_4}&{C_4}/{A_4}&-{D_4}/{A_4}&{1}/{A_4}
      \end{pmatrix}
 \begin{pmatrix}1\\k^-_{i}\\(k^-_{i})^2\\(k^-_{i})^3+{k_0^2}\,(V_{+}-V_{-})\left\{\omega-(V_++V_-)k^-_{i}\right\}
 \end{pmatrix}
\end{align}
with
\begin{align}
        &A_i=(k^+_i-k^+_j)(k^+_i-k^+_k)(k^+_i-k^+_l),\quad
        B_i=k^+_j\,k^+_k\,k^+_l,\\
        &C_i=k^+_j\,k^+_k+k^+_k\,k^+_l+k^+_l\,k^+_j,\quad
        D_i=k^+_j+k^+_k+k^+_l.\\
        &(\text{indices }i,j,k,l\text{ are  different each other}). \notag
\end{align}

\begin{figure}[H]
\centering
 \includegraphics[width=1\linewidth]{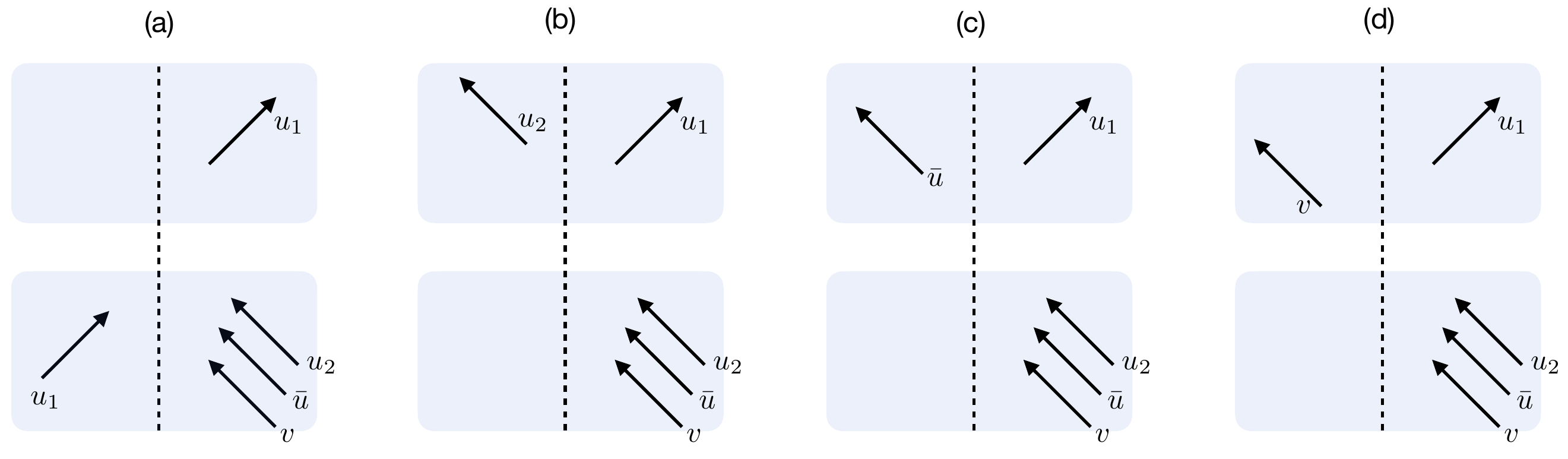}   
 \caption{Four different boundary conditions in $x<0$ for the wave
   equation~\eqref{eq:weq2} to determine the Bogoliubov
   coefficients. (a) Boundary condition with
   $\phi^{-}(x)=\phi_{u_1}^\text{in}$. For the subsonic case with
   $\omega_\text{GVH}<\omega$ or the transsonic case, the mode $u_1$
    becomes decaying mode. (b) Boundary condition with
   $\phi^{-}(x)=\phi_{u_2}^\text{out}$. (c) Boundary condition with
   $\phi^{-}(x)=\phi_{\bar{u}}^\text{out}$. (d) Boundary condition
   with $\phi^{-}(x)=\phi_{v}^\text{out}$.}
 \label{fig:vout}
\end{figure}

By specifying a mode in $x<0$, it is possible to obtain a wave
function which satisfies a given boundary condition in
$x<0$. Schematic diagrams describing possible four different boundary
conditions in $x<0$ region are shown in Fig.~\ref{fig:vout}.
\noindent
For the plane wave $\exp(i k_i^- x)$ in $x<0$ with a real wave number
$k_i^-$, we can define the normalized mode function
$\phi^\text{in/out}_i(x)=\exp(i\, k_i^\text{in/out}\, x)/N^{\pm}_i$
with
\begin{equation}
   \label{norm}
        N^{\pm}_i=\sqrt{4\pi c_{s}(k_{i}^{\pm})\, k_{i}^{\pm}\,v_{g}(k_{i}^{\pm})},\quad  v_{g}=\left(\frac{dk}{d\omega}\right)^{-1}.
\end{equation}
We put labels ``$\pm$'' and ``in/out'' depending on asymptotic regions
and the sign of the group velocity. If all modes are normalizable
(i.e. all solutions of the dispersion relation are real), plane wave
solutions with specified boundary conditions are given as follows:
\begin{align}
        \label{eq:u1in}
        &(\text{a})\quad\phi(x)=
        \begin{cases}
          N^-_{u_1}\,\phi^\text{in}_{u_1} & (x<0)\\
          N^+_{u_1}\,\alpha_{u_1 u_1}\,\phi^\text{out}_{u_1}
                        +N^+_{u_2}\,\alpha_{u_1 u_2}\,\phi^\text{in}_{u_2}
                        +N^+_{\bar{u}}\,\alpha_{u_1 \bar{u}}\,\phi^\text{in}_{\bar{u}}
                        +N^+_{v}\,\alpha_{u_1 v}\,\phi^\text{in}_{v} & (x>0)
        \end{cases}
         \\
       \label{eq:u2out}
        &(\text{b})\quad\phi(x)=
                \begin{cases}
                        N^-_{u_2}\,\phi^\text{out}_{u_2} & (x<0)\\
                        N^+_{u_1}\,\alpha_{u_2 u_1}\,\phi^\text{out}_{u_1}
                        +N^+_{u_2}\,\alpha_{u_2 u_2}\,\phi^\text{in}_{u_2}
                        +N^+_{\bar{u}}\,\alpha_{u_2 \bar{u}}\,\phi^\text{in}_{\bar{u}}
                        +N^+_{v}\,\alpha_{u_2 v}\,\phi^\text{in}_{v} & (x>0)\\
        \end{cases}
        \\
        \label{eq:ubarout}
        &(\text{c})\quad\phi(x)=
                \begin{cases}
                        N^-_{\bar{u}}\,\phi^\text{out}_{\bar{u}} & (x<0)\\
                        N^+_{u_1}\,\alpha_{\bar{u} u_1}\,\phi^\text{out}_{u_1}
                        +N^+_{u_2}\,\alpha_{\bar{u} u_2}\,\phi^\text{in}_{u_2}
                        +N^+_{\bar{u}}\,\alpha_{\bar{u} \bar{u}}\,\phi^\text{in}_{\bar{u}}
                        +N^+_{v}\,\alpha_{u_1 v}\,\phi^\text{in}_{v} & (x>0)\\
        \end{cases}
        \\
        \label{eq:vout}
        &(\text{d})\quad\phi(x)=
                \begin{cases}
                        N^-_{u_1}\,\phi^\text{out}_{u_1} & (x<0)\\
                        N^+_{u_1}\,\alpha_{v u_1}\,\phi^\text{out}_{u_1}
                        +N^+_{u_2}\,\alpha_{v u_2}\,\phi^\text{in}_{u_2}
                        +N^+_{\bar{u}}\,\alpha_{v \bar{u}}\,\phi^\text{in}_{\bar{u}}
                        +N^+_{v}\,\alpha_{v v}\,\phi^\text{in}_{v} & (x>0)\\
        \end{cases}.
\end{align}
Even if there exists unnormalizable modes, the logic is essentially
same, but we have to treat the norm of modes more carefully.  From
Eqs.~\eqref{eq:u1in}-\eqref{eq:vout}, we can read off relations
between the in-mode functions and the out-mode functions.  For
example, let us consider Eq.~(\ref{eq:u1in}).  In the asymptotic out
region, the wave function is expressed as
$\phi(x)=N^+_{u_1}\alpha_{u_1 u_1}\phi^\text{out}_{u_1}$, thus this
mode defines the out-vacuum state.  In the asymptotic in region, the
wave function is expressed as superposition of plane waves
\begin{align}
        \phi(x)=N^-_{u_1}\,\phi^\text{in}_{u_1}+N^+_{u_2}\,\alpha_{u_1 u_2}\,\phi^\text{in}_{u_2}+N^+_{\bar{u}}\,\alpha_{u_1 \bar{u}}\,\phi^\text{in}_{\bar{u}}+N^+_{v}\,\alpha_{u_1 v}\,\phi^\text{in}_{v}.
\end{align}
Therefore, we obtain the following in-out relation
\begin{align}
        N^+_{u_1}\,\alpha_{u_1 u_1}\,\phi^\text{out}_{u_1}=N^-_{u_1}\,\phi^\text{in}_{u_1}+N^+_{u_2}\,\alpha_{u_1 u_2}\,\phi^\text{in}_{u_2}+N^+_{\bar{u}}\,\alpha_{u_1 \bar{u}}\,\phi^\text{in}_{\bar{u}}+N^+_{v}\,\alpha_{u_1 v}\,\phi^\text{in}_{v}.
\end{align}
Repeating the same procedure for other three boundary conditions, we
obtain other three in-out relations:
\begin{align}
&N_{u_2}^-\,\phi_{u_2}^\text{out}+N_{u_1}^+\,\alpha_{u_2u_2}\,\phi_{u_1}^\text{out}=N_{u_2}^+\,\alpha_{u_2u_2}\,\phi_{u_2}^\text{in}+N_{\bar u}^+\,\alpha_{u_2\bar u}\,\phi_{\bar u}^\text{in}+N_v^+\,\alpha_{u_2 v}\,\phi_v^\text{in},\\
&N_{\bar u}^-\,\phi_{\bar u}^\text{out}+N_{u_1}^+\,\alpha_{\bar u u_1}\,\phi_{u_1}^\text{out}=
N_{u_2}^+\,\alpha_{\bar u u_2}\,\phi_{u_2}^\text{in}+N_{\bar u}^+\,\alpha_{\bar u\bar u}\,\phi_{\bar u}^\text{in}+N_v^+\,\alpha_{u_1v}\,\phi_v^\text{in},\\
&N_{u_1}^-\,\phi_{u_1}^\text{out}+N_{u_1}^+\,\alpha_{v u_1}\,\phi_{u_1}^\text{out}=N_{u_2}\,\alpha_{vu_2}\,\phi_{u_2}^\text{in}+N_{\bar u}^+\,\alpha_{v\bar u}\,\phi_{\bar u}^\text{in}+N_v^+\,\alpha_{vv}\,\phi_v^\text{in}.
\end{align}
By taking the Klein-Gordon inner product with the field operator
$\hat{\phi}$ both sides, from Eq.~\eqref{eq:ainout}, we obtain the
transformation between the in-mode operators and the out-mode
operators. This transformation is the Bogoliubov transformation, and
the transformation is determined by $4\times4$ Bogoliubov
coefficients.  We obtain the following form of Bogoliubov
transformation
\begin{equation}
\begin{pmatrix}
        -\left(\hat{a}_{\bar{u}}^{\text{out}}\right)^{\dagger} \\
\hat{a}_{v}^{\text{out}} \\
\hat{a}_{u_{1}}^{\text{out}} \\
\hat{a}_{u_{2}}^{\text{out}}
\end{pmatrix}
=\begin{pmatrix}
        \tilde{\beta}_{\bar{u} \bar{u}} & \tilde{\beta}_{\bar{u} v} & \tilde{\beta}_{\bar{u} u_{1}} & \tilde{\beta}_{\bar{u} u_{2}} \\
        \tilde{\beta}_{v \bar{u}} & \tilde{\beta}_{u v} & \tilde{\beta}_{v u_{1}} & \tilde{\beta}_{v u_{2}} \\
        \tilde{\beta}_{u_{1} \bar{u}} & \tilde{\beta}_{u, v} & \tilde{\beta}_{u, u_{1}} & \tilde{\beta}_{u_{1} u_{2}} \\
        \tilde{\beta}_{u_{2} \bar{u}} & \tilde{\beta}_{u_{2} v} & \tilde{\beta}_{u_{2} u_{1}} & \tilde{\beta}_{u_{2} u_{2}}
\end{pmatrix}
\begin{pmatrix}
-\left(\hat{a}_{\bar{u}}^{\text{in}}\right)^{\dagger} \\
\hat{a}_{v}^{\text{in}} \\
\hat{a}_{u_{1}}^{\text{in}} \\
\hat{a}_{u_{2}}^{\text{in}}
\end{pmatrix},
\end{equation}
where coefficients are given by
\begin{equation}
        \widetilde{\beta}_{u_1 i}=\begin{cases}
\dfrac{N_{i}^{+}\, \alpha_{u_1 i}}{N_{u_{1}}^{+} \,\alpha_{u_{1} u_{1}}} & \quad\left(i \neq u_{1}\right) \\
        \dfrac{N^{-}_{u_{1}}}{N_{u_{1}}^{+}\, \alpha_{u_1 u_{1}}} &\quad \left(i=u_{1}\right)
\end{cases}
\end{equation}
and
\begin{equation}
        \widetilde{\beta}_{i j}=\begin{cases}
\dfrac{N_{j}^{+}\,\alpha_{i j}-N_{u_1}^{+}\,\alpha_{i u_1}\,\tilde{\beta}_{u_1 j}}{N_{i}^{-}}&\quad\left(j \neq u_{1}\right) \\[10pt]
-\dfrac{N_{u_1}^{+}\,\alpha_{i u_{1}}\,\tilde{\beta}_{u_1 u_{1}}}{N_{i}}& \quad\left(j=u_{1}\right)
\end{cases} .
\end{equation}
We obtain the Bogoliubov transformation which transforms the out-mode
operators to the in-mode operators:
\begin{equation}
\begin{pmatrix}
        \left(\hat{a}_{\bar{u}}^{\text{in}}\right)^{\dagger} \\
\hat{a}_{v}^{\text{in}} \\
\hat{a}_{u_{1}}^{\text{in}} \\
\hat{a}_{u_{2}}^{\text{in}}
\end{pmatrix}
=\begin{pmatrix}
\beta_{\bar{u} \bar{u}} & \beta_{\bar{u} v} & \beta_{\bar{u} u_{1}} & \beta_{\bar{u} u_{2}} \\
\beta_{v \bar{u}} & \beta_{u v} & \beta_{v u_{1}} & \beta_{v u_{2}} \\
\beta_{u_{1} \bar{u}} & \beta_{u_1 v} & \beta_{u_1 u_{1}} & \beta_{u_{1} u_{2}} \\
\beta_{u_{2} \bar{u}} & \beta_{u_{2} v} & \beta_{u_{2} u_{1}} & \beta_{u_{2} u_{2}}
\end{pmatrix}
\begin{pmatrix}
\left(\hat{a}_{\bar{u}}^{\text{out}}\right)^{\dagger} \\
\hat{a}_{v}^{\text{out}} \\
\hat{a}_{u_{1}}^{\text{out}} \\
\hat{a}_{u_{2}}^{\text{out}}
\end{pmatrix}.
\label{eq:Bogoin}
\end{equation}
where $\beta_{ij}$ satisfies the relation $\sum_k\tilde{\beta}_{ik}\beta_{kj}=\delta_{ij}$.
\section{Vacuum state and covariance matrix}

The Bogoliubov transformation related to the $\bar{u}$-mode is given by
\begin{equation}
  \hat{a}_{\bar{u}}^\text{in}=\beta^*_{\bar{u} \bar{u}}\,
  \hat{a}_{\bar{u}}^\text{out}+\beta^*_{\bar{u} v} (\hat{a}_{v}^\text{out})^\dag+\beta^*_{\bar{u} u_1} (\hat{a}_{u_{1}}^\text{out})^\dag+\beta^*_{\bar{u} u_{2}} (\hat{a}_{u_{2}}^\text{out})^\dag,
\end{equation}
where
$|\beta_{\bar{u} \bar{u}}|^2-|\beta_{\bar{u} v}|^2-|\beta_{\bar{u}
  u_1}|^2-|\beta_{\bar{u} u_2}|^2=1$ holds.  The equality
$\beta_{\bar{u} u_{2}}=0$ holds for the subsonic case with the GVH
and the transsonic case because there exist no $u_2$-mode in the
out-state.  The Bogoliubov coefficients related to $\bar{u}$-mode can
be parameterized as
\begin{align}
        &\beta_{\bar{u}\bar{u}}=e^{i\phi_1}\cosh r,\quad
        \beta_{\bar{u}v}=e^{i\phi_2}\sinh r\sin\theta, \notag\\
        &\beta_{\bar{u}u_1}=e^{i\phi_3}\sinh r\cos\theta\sin\phi,\quad
        \beta_{\bar{u}u_2}=e^{i\phi_4}\sinh r\cos\theta\cos\phi,
        \label{eq:rtp}
\end{align}
where $r,\theta,\phi,\phi_1,\phi_2,\phi_3,\phi_4$ are real parameters.
$r$ is the squeezing parameter and as $r\to 0$, the number of created
particles decreases.  $\theta$ represents ratio of the $u$-mode and
the $v$-mode; as $\theta\to0$, mixing of  $u$-mode and  $v$-mode
becomes small.  The parameter $\phi$ represents ratio of 
$u_1$-mode and  $u_2$-mode.


With these parameters, we can characterize the out-vacuum state.
Let us define new annihilation operators ${\hat{A}_1,\hat{A}_2,\hat{A}_3,\hat{A}_4}$ from  the in-mode annihilation operators by
\begin{align}
        \hat{A}_1&=\hat{a}^\text{out}_{\bar{u}},\\
        \hat{A}_2&=e^{i(\phi_3-\phi_1)}\sin\theta\cos\phi\ \hat{a}^\text{out}_{u_1}+e^{i(\phi_4-\phi_1)}\sin\theta\sin\phi\ \hat{a}^\text{out}_{u_2}+e^{i(\phi_2-\phi_1)}\cos\theta\ \hat{a}^\text{out}_v,\\
        \hat{A}_3&=-e^{i(\phi_3-\phi_1)}\cos\theta\cos\phi\ \hat{a}^\text{out}_{u_1}-e^{i(\phi_4-\phi_1)}\cos\theta\sin\phi\ \hat{a}^\text{out}_{u_2}+e^{i(\phi_2-\phi_1)}\sin\theta\ \hat{a}^\text{out}_v,\\
        \hat{A}_4&=-e^{i(\phi_3-\phi_1)}\sin\phi\ \hat{a}^\text{out}_{u_1}+e^{i(\phi_4-\phi_1)}\cos\phi\ \hat{a}^\text{out}_{u_2}.
\end{align}
These new operators satisfy $[\hat{A}_i,\hat{A}_j]=0$ and
$[\hat{A}_i,\hat{A}_j^\dag]=\delta_{ij}$.  With these new operators,
from Eq.~\eqref{eq:Bogoin}, annihilation operators of the in-mode can
be written as
\begin{align}
        &\hat{a}^\text{in}_{\bar{u}}=e^{i\phi_1}(\cosh r \, \hat{A}_1+ \sinh r \, \hat{A}_2^\dag),\\
        &\hat{a}^\text{in}_{v}= e^{i\phi_1'}\left[\frac{\rho}{\cosh r}(\sinh r \,\hat{A}_1^\dag+ \cosh r \, \hat{A}_2)+\sqrt{1-\frac{|\rho|^2}{\cosh^2 r}}\,(\cos\phi'\, \hat{A}_3+\sin\phi'\, \hat{A}_4)\right],\\
        &\hat{a}^\text{in}_{u_1}= e^{i\phi_2'}\left[\frac{\rho'}{\cosh r}(\sinh r \, \hat{A}_1^\dag+ \cosh r \, \hat{A}_2)+\sqrt{1-\frac{|\rho'|^2}{\cosh^2 r}}\,(\cos\phi''\, \hat{A}_3+\sin\phi''\, \hat{A}_4)\right],\\
        &\hat{a}^\text{in}_{u_2}= e^{i\phi_3'}\left[\frac{\rho''}{\cosh r}(\sinh r \, \hat{A}_1^\dag+ \cosh r \, \hat{A}_2)+\sqrt{1-\frac{|\rho''|^2}{\cosh^2 r}}\,(\cos\phi'''\, \hat{A}_3+\sin\phi'''\, \hat{A}_4)\right],
\end{align}
where we introduced new constants
$\phi'_{1}, \phi'_{2}, \phi'_{3}, \phi'', \phi''', \rho, \rho',
\rho''$ which are related to original parameters
$\phi_{1}, \phi_{2}, \phi_{3}, \phi_{4},\beta_{ij}$.  From these
relations, the vacuum condition for the in-state yields
\begin{align}
&(\cosh r\,\hat A_1+\sinh r\,\hat A_2^\dag)\ket{0_\text{in}}=0,\quad (\sinh r\,\hat A_1^\dag+\cosh r\,\hat A_2)\ket{0_\text{in}}=0,\\
&\hat A_3\ket{0_\text{in}}=0,\quad \hat A_4\ket{0_\text{in}}=0,
\end{align}
and the in-vacuum state
is written as
\begin{equation}
        |0_\text{in}\rangle=\frac{1}{\cosh r}\sum_{n=0}^\infty(-\tanh r)^n |n_{A_1}\rangle|n_{A_2}\rangle|0_{A_3}\rangle|0_{A_4}\rangle.
\end{equation}
Thus the in-vacuum state is the two mode squeezed state of 
$A_1$-mode and  $A_2$-mode.

To quantify entanglement between each mode using the negativity, we
introduce canonical variables ${\hat{X}_i,\hat{P}_i}$ by
\begin{align}
        \hat{X}_i=\frac{\hat{a}_i^\text{in}+(\hat{a}_i^\text{in})^\dag}{\sqrt{2}},
        \quad\hat{P}_i=\frac{\hat{a}_i^\text{in}-(\hat{a}_i^\text{in})^\dag}{i\sqrt{2}},\quad
        [\hat X_i, \hat P_j]=i\,\delta_{ij}.
\end{align}
Then the wave function of the in-vacuum state is given as
\begin{align}
        \psi_0(X_1,X_2,X_3,X_4)&=\langle X_1,X_2,X_3,X_4|0_\text{in}\rangle \notag\\
        &=\frac{1}{\pi}\exp\left(-\frac{X_1^2+X_2^2+X_3^2+X_4^2}{2}\right).
        \label{eq:wavefunc}
\end{align}
The Wigner function of this wave function is defined by
\begin{align}
        W(\bs{X},\bs{P})&:=\frac{1}{(2\pi)^4}\int d^4\bs{Y}~e^{i\bs{P}\cdot \bs{Y}}\psi_0\left(\bs{X}-\frac{\bs{Y}}{2}\right)\psi_0^*\left(\bs{X}+\frac{\bs{Y}}{2}\right) \notag \\
        &=\frac{1}{\pi^3}\exp(-\bs{X}^2-\bs{P}^2).
\end{align}
Introducing a vector with canonical variables $\hat{\bs{\xi}}=(\hat{X}_1,\hat{P}_1,\hat{X}_2,\hat{P}_2,\hat{X}_3,\hat{P}_3,\hat{X}_4,\hat{P}_4)^T$, the covariance matrix is defined by
\begin{align}
        V_{ij}:=\left\langle\frac{\hat{\xi}_i\,\hat{\xi}_j+\hat{\xi}_j\,\hat{\xi}_i}{2}\right\rangle
        =\int d^8\bs{\xi} ~\xi_i\,\xi_j \,W(\bs{\xi}),
\end{align}
and for the wave function Eq.~(\ref{eq:wavefunc}),
$V_{ij}=\delta_{ij}/2$.  Since the Bogoliubov transformation preserves
commutation relations of creation and annihilation operators, it also
keeps commutation relations between canonical variables defined in
terms of creation and annihilation operators.

Now we introduce canonical variables for the out-modes as
\begin{align*}
        \hat{x}_i=\frac{\hat{a}_i^\text{out}+(\hat{a}_i^\text{out})^\dag}{\sqrt{2}},
        \quad\hat{p}_i=\frac{\hat{a}_i^\text{out}-(\hat{a}_i^\text{out})^\dag}{i\sqrt{2}},
\end{align*}
and introduce a vector with canonical variables for the in-mode as $\hat{\bs{\xi}}'=(\hat{x}_1,\hat{p}_1,\hat{x}_2,\hat{p}_2,\hat{x}_3,\hat{p}_3,\hat{x}_4,\hat{p}_4)^T$.
The relation between in and out canonical variables is given by
\begin{equation}
        \hat{\xi}_i=\sum_jS_i^{~j}\,\hat{\xi}'_j
\end{equation}
where
\begin{equation}
        S=
        \begin{pmatrix}
                ~\Re[\beta_{11}]&~\Im[\beta_{11}]&~\Re[\beta_{12}]&-\Im[\beta_{12}]&~\Re[\beta_{13}]&-\Im[\beta_{13}]&~\Re[\beta_{14}]&-\Im[\beta_{14}]\\
                ~\Im[\beta_{11}]&~\Re[\beta_{11}]&~\Im[\beta_{12}]&-\Re[\beta_{12}]&~\Im[\beta_{13}]&-\Re[\beta_{13}]&\Im[\beta_{14}]&-\Re[\beta_{14}]\\
                ~\Re[\beta_{21}]&~\Im[\beta_{21}]&~\Re[\beta_{22}]&-\Im[\beta_{22}]&~\Re[\beta_{23}]&-\Im[\beta_{23}]&~\Re[\beta_{24}]&-\Im[\beta_{24}]\\
                               ~\Im[\beta_{21}]&~\Re[\beta_{21}]&~\Im[\beta_{22}]&~\Re[\beta_{22}]&~\Im[\beta_{23}]&~\Re[\beta_{23}]&~\Im[\beta_{24}]&~\Re[\beta_{24}]\\
                ~\Re[\beta_{31}]&~\Im[\beta_{31}]&~\Re[\beta_{32}]&-\Im[\beta_{32}]&~\Re[\beta_{33}]&-\Im[\beta_{33}]&~\Re[\beta_{34}]&-\Im[\beta_{34}]\\
                ~\Im[\beta_{31}]&-\Re[\beta_{31}]&~\Im[\beta_{32}]&~\Re[\beta_{32}]&~\Im[\beta_{33}]&~\Re[\beta_{33}]&~\Im[\beta_{34}]&~\Re[\beta_{34}]\\
                ~\Re[\beta_{41}]&~\Im[\beta_{41}]&~\Re[\beta_{42}]&-\Im[\beta_{42}]&~\Re[\beta_{43}]&-\Im[\beta_{43}]&~\Re[\beta_{44}]&-\Im[\beta_{44}]\\
                ~\Im[\beta_{41}]&-\Re[\beta_{41}]&~\Im[\beta_{42}]&~\Re[\beta_{42}]&~\Im[\beta_{43}]&~\Re[\beta_{43}]&~\Im[\beta_{44}]&~\Re[\beta_{44}]
        \end{pmatrix}
\end{equation}
with the subscript of $\beta_{ij}$ represents $1,2,3,4=\bar{u},v,u_1,u_2$.
Since this transformation keeps canonical commutation relations,
the matrix $S$ satisfies
\begin{align*}
        S\,\Omega\, S^T=\Omega,\quad
        \Omega=\bigoplus_{i=1}^4
        \begin{pmatrix}
                0&1\\
                -1&0\\
        \end{pmatrix}.
\end{align*}
From $|\det S|=1$, the relation between the in-mode covariance matrix
$V$ and the out-mode covariance matrix $V'$ is derived as
\begin{align}
        V_{ij}&=\int d^8\bs{\xi}~\xi_i\xi_j\, W(\xi)
        =\sum_{k,l}\int d^8\bs{\xi}'\, S_i^{~k}S_j^{~l}\,\xi'_k\,\xi'_l\, W(S(\bs{\xi}')) \notag\\
        &=\left( S\,V'\,S^T \right)_{ij},
\end{align}
where $W(S(\bs{\xi}'))=W'(\bs{\xi}')$ is the Wigner function for $\bs{\xi}'$. 
Thus the covariance matrix for the out-mode can be written as
\begin{align}
        V'=S^{-1}V(S^T)^{-1}
        &=\begin{pmatrix}
                V_1&V_2&V_3&V_4\\
                *  &V_5&V_6&V_7\\
                *  &*&V_{8}&V_{9}\\
                *  &*  &*  &V_{10}\\
        \end{pmatrix},
\end{align}
where $V_{j}, j=1, \cdots,10$ denotes 2$\times$2 submatrices of
8$\times$8 covariance matrix $V'$.  For the Gaussian state considering
here, it is easy to obtain the covariance matrix for the reduced three
mode state by simply integrating out one mode:
\begin{align*}
        \widetilde V_{ij}&=\int d^6\bs{\xi}~\xi_i\,\xi_j\, \widetilde W(\bs{\xi})
      =\int d^6\bs{\xi}~\xi_i\,\xi_j \left(\int d^{2}\bs{\xi}\,{W}(\bs{\xi})\right)
        ={V}_{ij},
\end{align*}
where $\widetilde{W}(\bs{\xi})$ is the Wigner function for reduced
state, and $\tilde{V}_{ij}$ is the covariance matrix of the reduced
state.  Similar argument can be applied to the covariance matrix of a
two modes state.

Using the covariance matrix, it is possible to evaluate entanglement
negativity which quantifies bipartite entanglement for a given
bipartition of the total system (see Appendix \ref{sec:ap-A} for its
definition).

\section{Results}
Our analysis is performed for the subsonic case and the transsonic
case. For the subsonic case, there is right-moving modes with low
frequency $\omega$ in $x<0$, however, there exists a critical
frequency $\omega_\text{GVH}$ such that there is no right-moving modes
with frequencies $\omega>\omega_\text{GVH}$ in $x<0$.  This means that
modes with sufficiently high frequency can feel the effective sonic
horizon (GVH) at $x=0$ even for the subsonic case.  For the transsonic
case, there is no right-moving mode in the super sonic region $x<0$,
and the point $x=0$ is the sonic horizon.

In our analysis, we adopt two sets of parameters
$\{k_0=100, V_+=-0.4, V_-=-0.6\}$ (subsonic case) and
$\{k_0=100, V_+=-0.75, V_-=-1.25\}$ (transsonic case).  Corresponding
to the cutoff wave number $k_{0}$, the cutoff frequency $\omega_c$ is
determined by $V_+$ and $k_{0}$.  The value of $\omega_\text{GVH}$ is
given by a point at which the line $\Omega=c_{s}(k)\,k$ is tangent to
$\Omega=\omega-V_{-}\,k$ in the dispersion diagram:
\begin{equation}
        \omega_\text{GVH}=\frac{k_0}{16}\left(3V_-+\sqrt{V_-^2+8}\right)\sqrt{8-2V_-^2+2V_-\sqrt{V_-^2+8}}.
\end{equation}
The value of $\omega_c$ is given by a point at which the line
$\Omega=c(k)\,k$ is tangent to $\Omega=\omega-V_{+}\,k$ in the
dispersion diagram:
\begin{equation}
        \omega_c=\frac{k_0}{16}\left(3V_++\sqrt{V_+^2+8}\right)\sqrt{8-2V_+^2+2V_+\sqrt{V_+^2+8}}.
\end{equation}
For the subsonic case,
$\omega_c/k_0=0.240,~ \omega_\text{GVH}/k_0=0.133$ and
$\omega_\text{GVH}/\omega_c=0.554$. For the transsonic case,
$\omega_c/k_0=0.0666$.

\subsection{Power spectrum of created particles}
We define the power spectrum of  out-going particles (radiations) as
\begin{equation}
       f_{u_{1}}(\omega)=|\beta_{u_{1}\bar u}|^{2},\quad f_{u_{2}}(\omega)=|\beta_{u_{2}\bar{u}}|^2,
       \quad f_{v}(\omega)=|\beta_{v\bar u}|^{2},\quad
        f_{\bar{u}}(\omega)=1-|\beta_{\bar{u}\bar{u}}|^2.
\end{equation}
For the power spectrum $f_i(\omega)$ of radiations, we introduce the
effective temperature $T_i(\omega)$ by the relation
\begin{equation}
  f_i(\omega)=\frac{1}{c_i\, e^{\omega/T_i(\omega)}-1},
  \label{eq:planck}
\end{equation}
where $c_i$ is a constant to be determined to fit the spectrum with Eq.~\eqref{eq:planck}.
If the effective temperature is constant with respect to $\omega$ in
some frequency range, the power spectrum has Planckian distribution  in that frequency range
and radiation cannot be distinguished from the thermal one.

\subsubsection{Subsonic case}
We obtain the analytical formula of the power spectrum for
$\omega/k_0\ll 1$.  For the subsonic case, the power spectrum is
\begin{equation}
|\beta_{u_1\bar{u}}|^2\sim\frac{\sqrt{1-V_+}\,
  (V_+-V_-)^2}{4(V_++1)^{3/2}
  (V_-+1)^2}\,\frac{\omega}{k_{0}},\quad\omega/k_0\ll 1.
\label{eq:pow-sub}
\end{equation}
Particle creation in low frequency region occurs due to the
Planckian mode associated with the non-linear dispersion. Actually for
$k_{0}\rightarrow\infty$ with fixed $\omega$, the created particle
number becomes zero. For high frequency region over
$\omega_{\text{GVH}}$, particle creation occurs due to the mode conversion
associated to the GVH, which is also related with the Planckian mode.

\begin{figure}[H]
        \centering
          \includegraphics[width=0.8\linewidth]{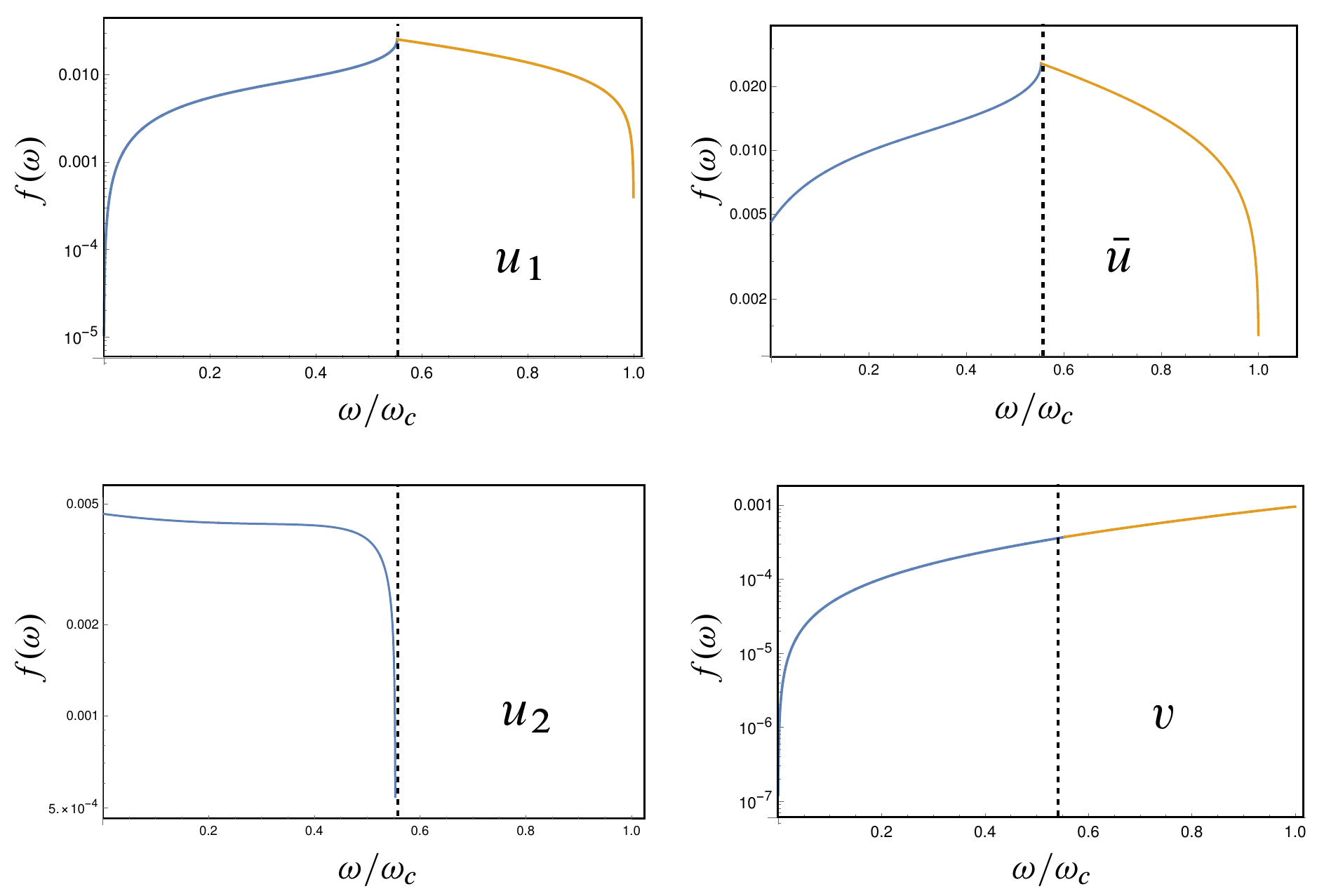}              
          \caption{Power spectrums of created particles for the
            subsonic case. $\omega_\text{GVH}/\omega_c=0.554$ for the
            present parameters. Across $\omega=\omega_\text{GVH}$, the
            number of modes changes from four to three and the
            spectrums of $u_1, \bar u, u_2$ are not smooth at $\omega_\text{GVH}$.  }
        \label{fig:spectrum}
\end{figure}

\begin{figure}[H]
        \centering  
      \includegraphics[width=0.4\linewidth]{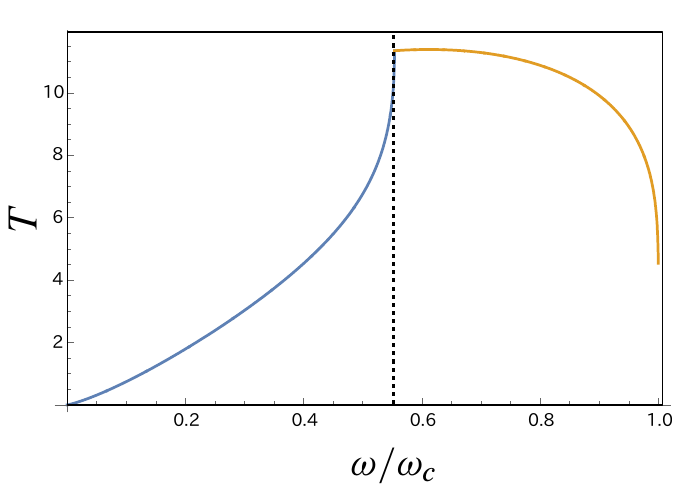}              
      \caption{Effective temperature of $u_1$-particle for the
        subsonic case.  Around $\omega\sim\omega_\text{GVH}$, the
        effective temperature becomes constant which indicates
        approximate thermal property of the radiation related to the GVH.  In this plot,
        we adopt $c_i=12.57$ in Eq.~\eqref{eq:planck}.}
        \label{fig:T1}
\end{figure}

We plot our result of power spectrums in Fig.~\ref{fig:spectrum}, where
frequency is normalized so as the cutoff frequency becomes equal to 1.
From $\omega$ dependence of the effective temperature
Fig.~\ref{fig:T1}, the thermality of the spectrum is not observed for
$\omega<\omega_\text{GVH}$.  In the $\omega\rightarrow 0$ limit, the
number of $u_1$-particle becomes zero, but 
finite numbers of $\bar{u}$-particle and $u_2$-particle are created,
and the numbers of these particles are almost same. However, the
behavior of power spectrums for these particles in the higher
frequency region is quite different. The number of $\bar{u}$-particle
increases with the increase of frequency until frequency reaches
$\omega_\text{GVH}$, at which the GVH appears. After the GVH is
formed, the number of $u_1, \bar u$- particles decreases as frequency
increases.  The number of $u_2$-particles decreases with the increase
of frequency, and becomes zero after the GVH is formed.  The number of
$v$-particle increases smoothly across $\omega=\omega_\text{GVH}$ as
frequency increases.
\noindent
Although there is no sonic horizon in this case, the effective
temperature becomes constant above $\omega_\text{GVH}$ and indicates
thermal property related to the GVH of emitted radiation around this
frequency region (Fig.~\ref{fig:T1}).

\subsubsection{Transsonic case}
\begin{figure}[H]
        \centering
         \includegraphics[width=1.0\linewidth]{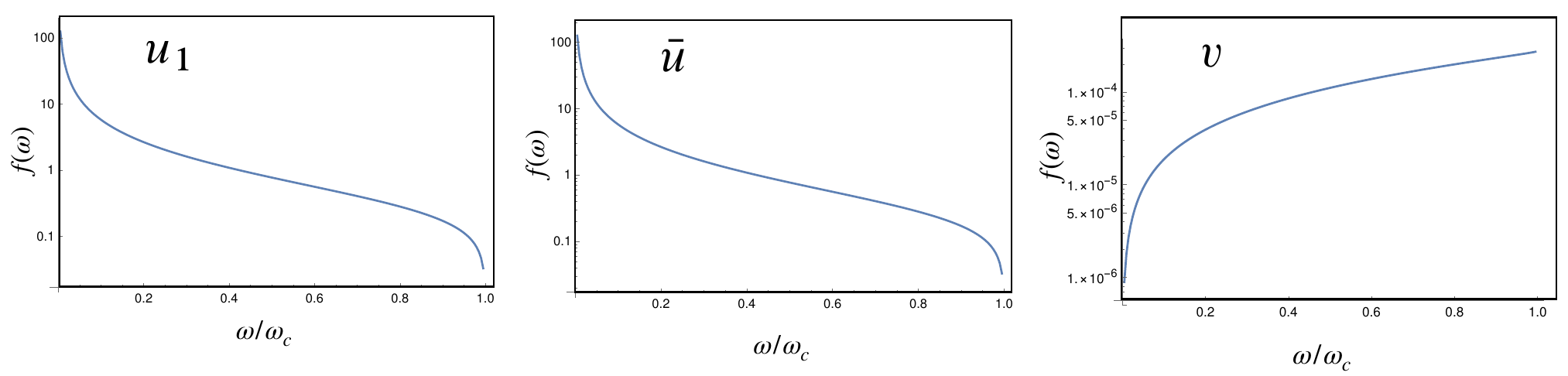}
             \caption{Power spectrums for the transsonic case. } 
        \label{fig:pnum2}
\end{figure}

\begin{figure}[H]
        \centering
                        \includegraphics[width=0.4\linewidth]{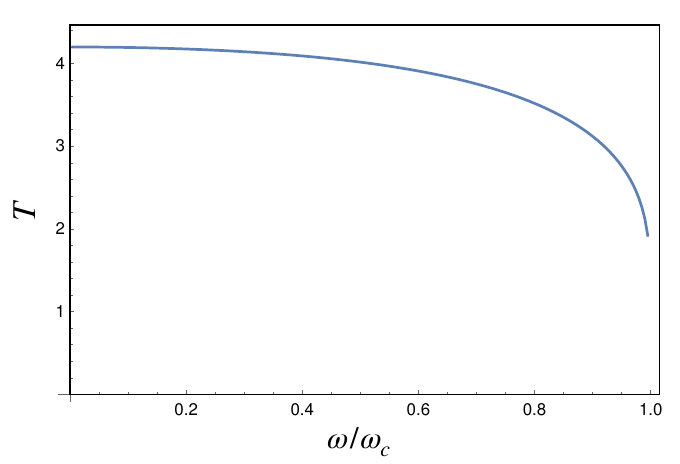}              
                        \caption{Effective temperature of 
                          $u_1$-particle for the transsonic
                          case. Around $\omega\sim 0$, the effective
                          temperature becomes constant. In this plot,
                          we adopt $c_i=1$ in Eq.~\eqref{eq:planck}. }
        \label{fig:T2}
\end{figure}

In this case, the power spectrum in the low frequency region is
\begin{equation}
        |\beta_{u_1\bar{u}}|^2\sim\frac{(V_++1)^{3/2} (V_-+1) (V_++V_-)}{\sqrt{1-V_+}\, (-V_-+1) (V_+-V_-)}\frac{k_{0}}{\omega}, \quad\omega/k_0\ll 1.
        \label{eq:pow-trans}
\end{equation}
Figure \ref{fig:pnum2} shows power spectrums of emitted radiations in
this case.  In the low frequency region, spectrums of
$u_1,\bar{u}$-particles are thermal and they decrease rapidly near the
cutoff frequency $\omega_{c}$.  The number of the $v$-particle shows
the similar behavior as that for the subsonic case.
\noindent
Figure \ref{fig:T2} shows behavior of the effective temperature for
 $u_1$-particle. Around $\omega\sim 0$, it becomes constant which
reflects the thermal property of emitted radiation.
\noindent
For the transsonic case with $\omega\sim 0$, neglecting
contribution of  $v$-mode, the Bogoliubov coefficients satisfy
$|\beta_{u_1u_2}|^2-|\beta_{u_1\bar u}|^2\approx 1$. These
coefficients diverge as $1/\omega$. Using this relation, the power
spectrum of $u_1$-particle is
\begin{equation}
 |\beta_{u_1\bar u}|^2\approx \left(\left|\frac{\beta_{u_1u_2}}{\beta_{u_1\bar u}}\right|^2-1\right)^{-1}.
\end{equation}
The ratio $|\beta_{u_1u_2}/\beta_{u_1\bar{u}}|$ determines the power
spectrum of  $u_1$-particle.  As $\omega\to0$ the ratio
$|\beta_{u_1u_2}/\beta_{u_1\bar{u}}|$ goes to 1 for the trans-sonic
case, this behavior of the the Bogoliubov coefficients is originated
from the boundary condition for the decaying wave function in $x<0$
(see Appendix \ref{sec:ap-B} for details).  The ratio can be
approximated as
\begin{equation}
    \left|\frac{\beta_{u_1u_2}}{\beta_{u_1\bar{u}}}\right|\simeq1+\gamma\,\omega\approx e^{\gamma\omega},
\end{equation}
where $\gamma$ is a factor determined by $V_\pm$.  This approximation
indicates that the power spectrum of  $u_1$-particle shows thermal
distribution with effective temperature $T=1/(2\gamma)$ in the low
frequency range. Using \eqref{eq:pow-trans}, the temperature is given
by
\begin{equation}
  T(\omega=0)=\frac{(V_++1)^{3/2} (V_-+1) (V_++V_-)}{\sqrt{1-V_+}\,
    (-V_-+1) (V_+-V_-)}\,k_{0}.
  \label{eq:efftemp}
\end{equation}
This formula provides a numerical value of the temperature as
$T(\omega=0)=4.20$ for present parameters and is consistent with the
numerical result (Fig.~\ref{fig:T2}).  However, this temperature seems
to be nothing to do with the surface gravity of the horizon because it
diverges for the step velocity profile, and the thermal property
appears due to non-linear dispersion relation (the Planckian
mode). Indeed, the temperature \eqref{eq:efftemp} can be regarded as
corresponding to the effective surface gravity which is defined by
velocity difference divided by the effective thickness of the sonic horizon determined
by the cutoff wave number $k_0$.

\subsection{Entanglement Structure}
\subsubsection{Subsonic case }
\begin{figure}[H]
 \centering
      \includegraphics[width=1\linewidth]{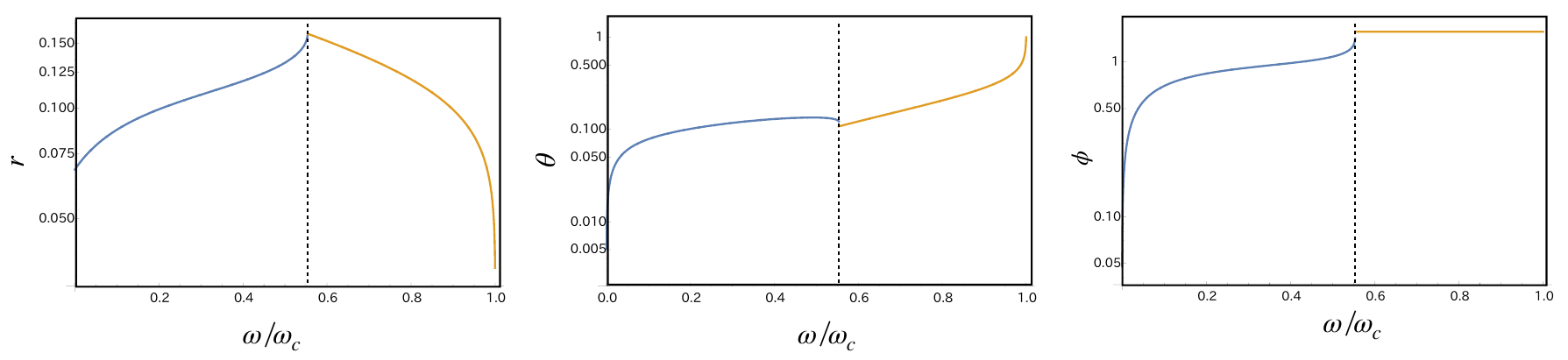}
      \caption{Behavior of parameters $r,\theta,\phi$ for the subsonic case.}
         \label{fig:rtp1}  
\end{figure}

To analyze entanglement between each particle mode, we calculated
parameters $r, \theta, \phi$ introduced in the previous section. These
parameters determine components of the covariance matrix for the
vacuum state.  Figure~\ref{fig:rtp1} shows behavior of these
parameters for the sub-sonic case. The squeezing parameter $r$
increases with the increase of frequency until $\omega_\text{GVH}$,
and then decreases with the increase of $\omega$.  The mixing
parameters $\theta$ and $\phi$ go to zero as $\omega\to0$, and
increase with the increase of $\omega$. Thus from the definition of
parameters \eqref{eq:rtp},  $u_{2}$-particle (Plankian mode) is
mainly created for $\omega\to0$. As $\omega$ increases, the number of
$u_{1}$-particle increases until $\omega_\text{GVH}$. For
$\omega_\text{GVH}<\omega$, as the GVH is formed, creation of 
$u_2$-particle is shut down and  $u_1$-particle and 
$\bar u$-particle mainly contribute as created particles.
  
\begin{figure}[H] 
 \centering
      \includegraphics[width=0.8\linewidth]{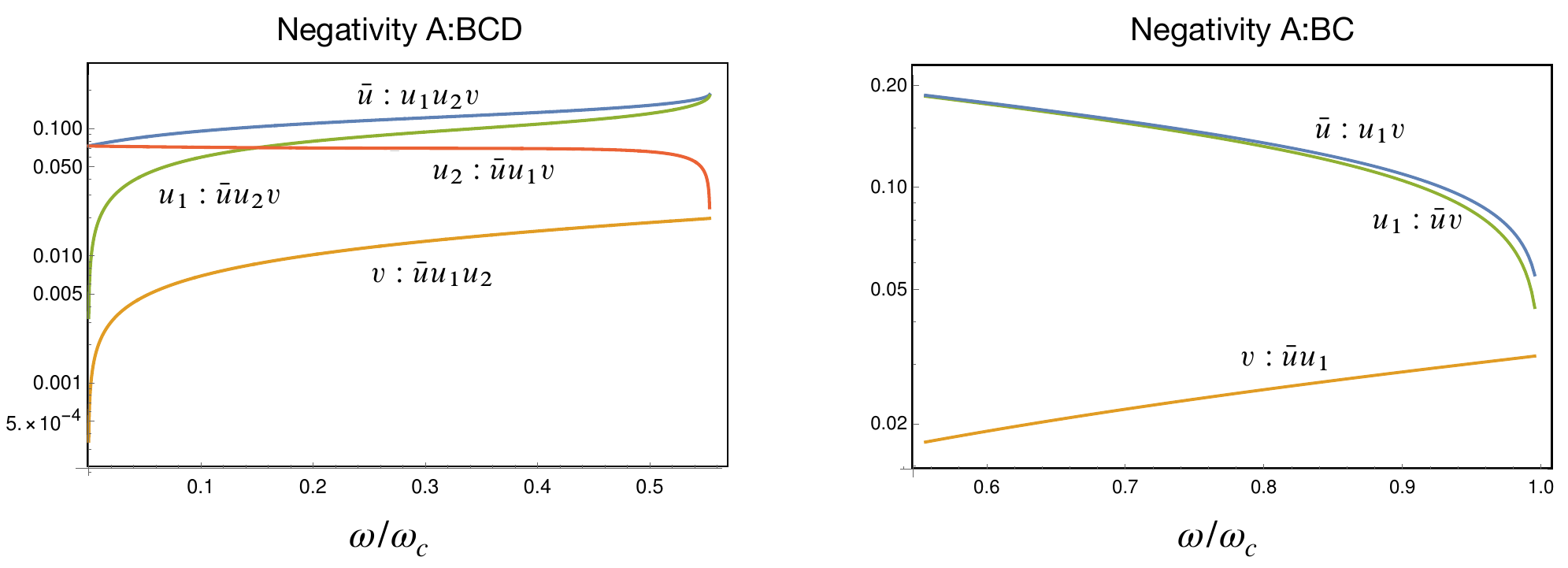}
      \caption{Negativity for the subsonic case. For low frequency
        $\omega<\omega_\text{GVH}$ (left panel), the number of
        particle modes is four. For high frequency
        $\omega_\text{GVH}<\omega$ (right panel), the number of
        particle modes is three.}
         \label{fig:neg1a}  
\end{figure}

\begin{figure}[H]
 \centering
      \includegraphics[width=1.0\linewidth]{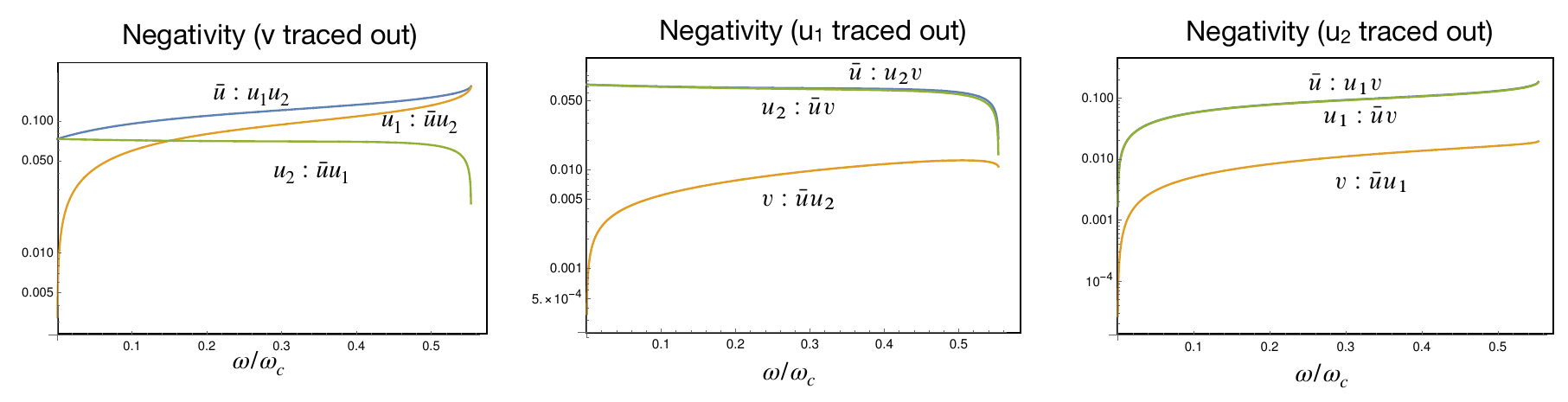}
      \caption{Negativity of reduced state for the subsonic case in
        the low frequency region $0<\omega<\omega_\text{GVH}$.}
         \label{fig:neg1b}  
\end{figure}

\begin{figure}[H]
 \centering
      \includegraphics[width=0.7\linewidth]{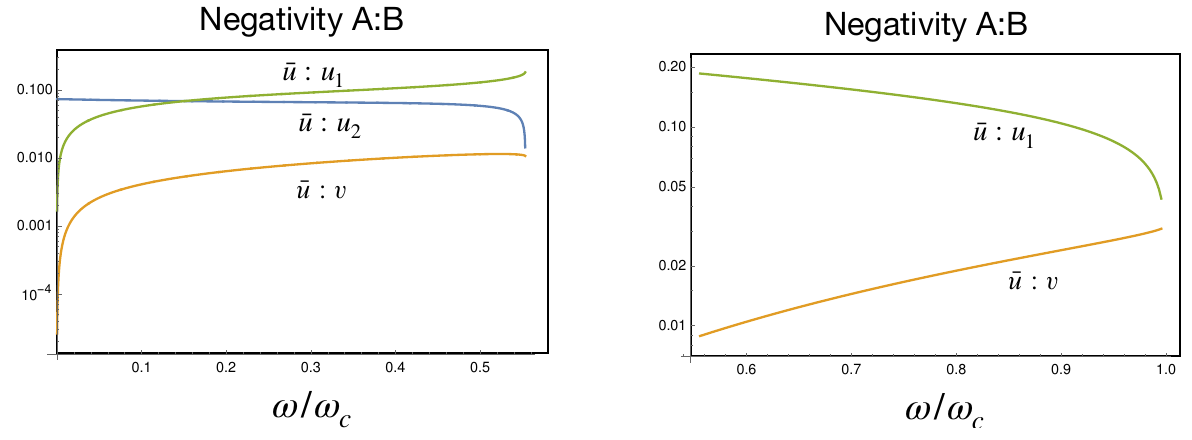}
      \caption{Negativity of reduced two mode state for the subsonic
        case.}
         \label{fig:neg1c}  
\end{figure}

Behavior of the entanglement negativity for the subsonic case is shown
in Fig.~\ref{fig:neg1a}, Fig.~\ref{fig:neg1b} and
Fig.~\ref{fig:neg1c}.  Figure \ref{fig:neg1a} is the negativity for
bi-partitioning of the total pure system (four modes for
$\omega<\omega_\text{GVH}$ (left panel) and three modes for
$\omega_\text{GVH}<\omega$ (right panel)).  For
$\omega<\omega_\text{GVH}$, entanglement between $u_1, v$-particles
and other particles goes to zero as $\omega\to0$. This decrease of
entanglement corresponds to the decrease of the created number of
$u_1$-particle and $v$-particle. Entanglement between
$\bar{u},v,u_1$-particles and other particles increases with the
increase of $\omega$, whereas entanglement between $u_2$-particle and
other particles decreases.  For $\omega_\text{GVH}<\omega$ where the
GVH exists, $u_2$-particle disappears and the total number of modes
becomes three. Entanglement between $\bar{u},u_1$-particles and other
particles decreases and entanglement between $v$-particle and other
particles increases with the increase of $\omega$.

\begin{figure}[H] 
 \centering
      \includegraphics[width=0.8\linewidth]{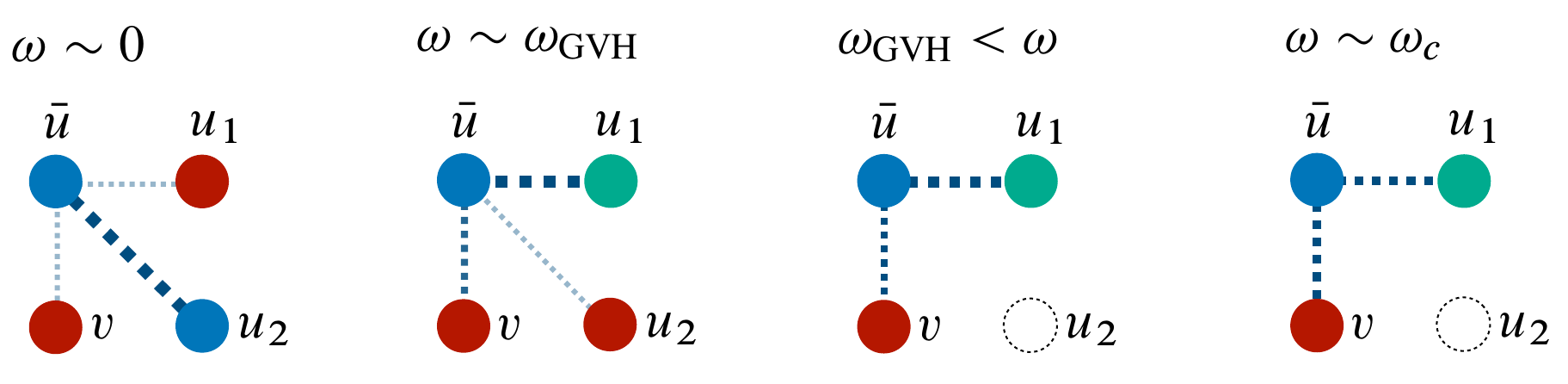}
      \caption{ Schematic pictures of entanglement
        structure for the subsonic case. Red disks represent
        non-Planckian modes, blue disks represent Planckian mode, and
        green disks represent sub-Planckian modes. For low frequency,
        entanglement of the system is shared mainly by $\bar u$-$u_2$
        pair. For $\omega\sim\omega_\text{GVH}$, entanglement of the
        system is shared mainly by $\bar u$-$u_1$ pair.}
         \label{fig:sche1}  
\end{figure}

In the limit of $\omega\rightarrow 0$, $u_1$ and $v$ modes are
separable from other three modes, and $u_2$ and $\bar{u}$ mode forms
an entangled pair.  With the increase of frequency, entanglement
between $u_1$ and $\bar{u}$ modes, and entanglement between $v$ and
$\bar{u}$ modes become larger. And near the frequency
$\omega_\text{GVH}$, entanglement between $u_1$ and $\bar{u}$ modes
becomes the main contribution to entanglement of the four modes
system. For $\omega>\omega_\text{GVH}$, entanglement between $u_1$ and
$\bar{u}$ modes starts to decrease, whereas entanglement between $v$
and $\bar{u}$ modes keeps increase and their amount become comparable
near the cutoff frequency $\omega_c$.  We present schematic pictures
of the entanglement structure in Fig.~\ref{fig:sche1}.
\noindent
For $\omega\sim0$,  non-Planckian modes $u_1, v$ can not entangle
with  Planckian modes $u_2, \bar u$.  With the increase of
frequency, the non-Planckian mode $u_1$ becomes the sub-Planckian mode,
and $\bar u$ and $u_1$ are entangled.  This is the reason why
entanglement between $u_1$ and other modes gets larger with the
increase of  frequency
 in the low frequency region.

Now let us comment on the thermal property of radiation for
  $\omega_\text{GVH}<\omega$ where the GVH exists. For models with
slowly varying velocity profiles, in the vicinity of the GVH, the wave
number corresponding to emitted particles is expressed as
\begin{align}
        k(x)=\frac{\omega-2\{c_s(k(x_0))+v(x_0)\}}{k(x_0)\,\kappa(x-x_0)-\{v(x_0)+c_s(k(x_0))\}}k(x_0),
\end{align}
where $x_0$ is the location of the GVH, $\kappa$ is the first
derivative of the velocity profile at the GVH, $k(x_0)$ is the wave
number at the GVH.  This $x$-dependent wave number leads to the
logarithmic behavior of the phase factor
$S(x,t)=\int^x dx'k(x')-\omega \,t$ with a branch point
$x'_0=x_0+(v(x_0)+c_s(k(x_0)))/k(x_0)\kappa$.  Let us introduce the
WKB mode function $\psi_\pm=\exp(iS(x,t))$ for $x\neq x'_0$. Then we
obtain the positive norm in-mode function $\psi_\text{in}(x,t)$ as
\begin{align}
        \label{eq:wkbin}
        \psi_\text{in}(x,t)\sim\psi_{+}(x,t)+\exp\left(\frac{-\pi\{\omega-2(c_s(k(x_0))+v(x_0))\}}{\kappa}\right)\psi_{-}(x,t),
\end{align}
which is obtained by requiring that $\psi_\text{in}(x,t)$ is analytic
in the upper half complex $x$-plane.  Equation~(\ref{eq:wkbin})
implies that the wave function includes the thermal-like factor but it
deviates from the Planckian distribution owing to the existence of
$c_s(k(x_0))+v(x_0)$ terms, which also have frequency dependence.  The
bipartite entanglement for $\omega_\text{GVH}<\omega$ decreases with
the increase of $\omega$, which is the same behavior for the
transsonic case and related to the thermal property of
radiations. These considerations suggest that the similar effect
appears for the steep velocity profile case and results in approximate
thermal behavior for $\omega_\text{GVH}<\omega$.

\subsubsection{Transsonic case}
\begin{figure}[H]
 \centering
      \includegraphics[width=1\linewidth]{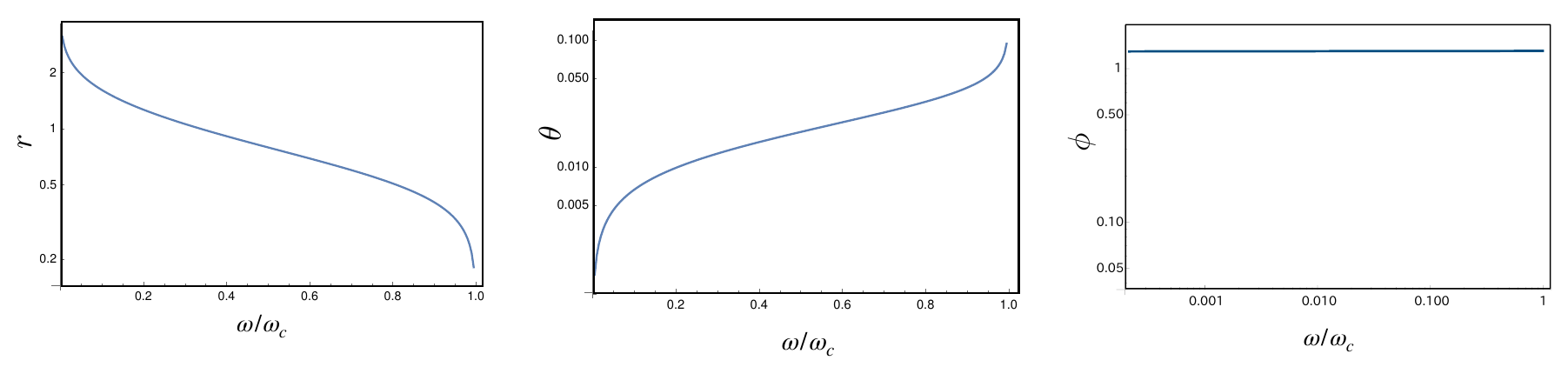}
       \caption{Frequency dependence of parameters $r,\theta, \phi$ for the transsonic case.}
         \label{fig:rtp2}  
\end{figure}

\begin{figure}[H] 
  \centering \includegraphics[width=0.7\linewidth]{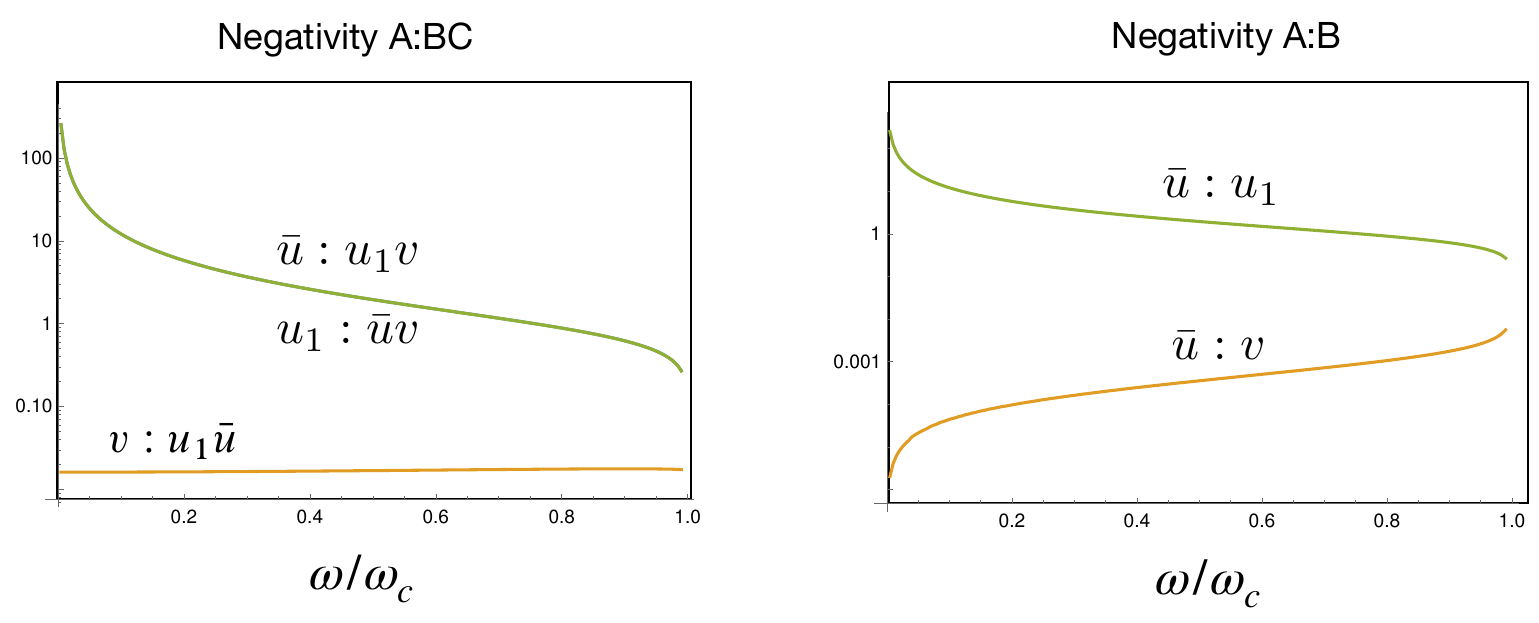}
      \caption{Behavior of negativity for the transsonic case.}
         \label{fig:neg2}  
\end{figure}

\begin{figure}[H]
 \centering
      \includegraphics[width=0.4\linewidth]{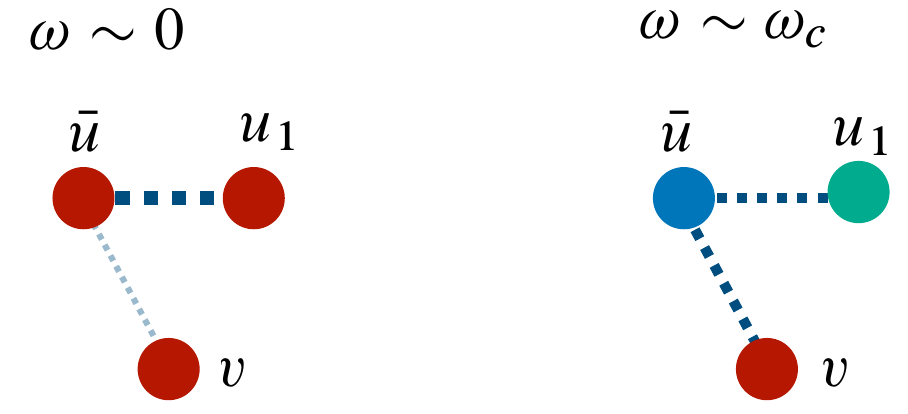}
      \caption{Schematic pictures of entanglement structure for the
        transsonic case. Red disks represent non-Planckian modes, blue
        disks represents Planckian modes, and green disks represents
        sub-Planckian modes. For low frequency, entanglement of the
        system is shared maily by $\bar u$-$u_1$ pair.}
         \label{fig:sche2}  
\end{figure}

Figure~\ref{fig:rtp2} shows $\omega$ dependence of parameters
$r,\theta,\phi$, and Fig.~\ref{fig:neg2} shows negativity for the
transsonic case.  Entanglement between $\bar{u}, u_1$ modes and
other modes decreases with the increase of $\omega$, and entanglement
between $v$-mode and other modes also increase with the
increase of $\omega$.  This behavior is consistent with that of the
power spectrum; with the increase of entanglement, the number of
created particles increases.
For $\omega\rightarrow 0$ limit, the $v$-mode becomes approximately separable from
other modes,  $u_1$ and $\bar{u}$ are entangled.  With the increase
of $\omega$, entanglement between $u_1$ and $\bar{u}$ modes decreases
and entanglement between $v$ and $\bar{u}$ modes increases. And the 
amount of these entanglement becomes comparable near the cutoff
frequency $\omega_c$. This behavior is same as that observed in our
previous study for the transsonic flow with finite surface gravity at
the sonic horizon~\cite{nambu}. Schematic structure of entanglement for
the transsonic case is shown in Fig.~\ref{fig:sche2}.

\section{Conclusion}
We have calculated the power spectrum and entanglement of the scalar
field modes in the dispersive media with a step velocity profile.  For
the transsonic case, we have obtained the similar result as
\cite{nambu}, but the temperature of the radiation is given by
Eq.~(\ref{eq:efftemp}), which is not equal to derivative of the fluid
velocity at the sonic horizon.  For the subsonic case, the situation
is completely different.  Entanglement between $u_1$-mode and
$\bar{u}$-mode, and the power spectrum of created $u_1$-particle
increases with frequency $\omega$ until the frequency
reaches $\omega_\text{GVH}$ where the GVH appears. The power spectrum
becomes a decreasing function of frequency for
$\omega_\text{GVH}<\omega$.  For the dispersive model investigated in
this paper, the power spectrum of $u_1$-mode for the subsonic
case and for the transsonic case have the similar behavior in the high
frequency region; it is not possible to distinguish them  if we only
measure the power spectrum for high energy
particles emitted from the step.  Concerning entanglement
  structure, we found that Planckian modes can not entangle with
  non-Planckian modes (see Fig.~\ref{fig:sche1} and
  Fig.~\ref{fig:sche2}); For the subsonic case, in the low frequency
limit, $u_1$-mode and $v$-mode are non-Planckian modes, and
$\bar{u}$-mode and $u_2$-mode are Planckian modes. Entanglement of the
system is shared only between $u_2$-mode and $\bar{u}$-mode, and the
$u_1$-particle is not created. With the increase of frequency,
$u_1$-mode becomes sub-Planckian mode and $u_1$-particle can be
created. For the transsonic case, all of the modes are sub-Planckian
modes in the low frequency limit, and $u_1$-mode and $\bar{u}$-mode
can entangle.

Although we did not treat in this paper, we are interested in the
following issues; the first one is how the cutoff scale affects the
total energy and the total entanglement of modes.  We do not
understand how the total energy of modes and entanglement shared
between modes with  non-linear dispersions.  The second one
is behavior of two point correlation functions.  Two point correlation
functions for analog black holes are investigated in
\cite{Schutzhold2,Steinhauer}.  It may be interesting to evaluate them
for the subsonic case without a sonic horizon.  The third one is
dependence of dispersion relation on particle creations and
entanglement.  We considered the subluminal dispersion in this paper,
but for the superluminal dispersion, the number of the negative norm
modes is different and we expect different entanglement structure.
These problems are left for our future research.

\acknowledgements{
Y.N. was supported in part by JSPS KAKENHI Grant No. 19K03866. }

\appendix
\section{Entanglement negativity}
\label{sec:ap-A}
Entanglement of the in-vacuum state is evaluated using the positive
partial transpose (PPT) criterion for continuous
variable~\cite{Peres,Horodecki,Simon}.  The PPT criterion states that
if a partially transposed density matrix has negative eigen values,
the bipartite state is entangled.  For bosonic systems, we can rewrite
the PPT criterion in terms of a covariance matrix. From positive
definiteness of the density matrix and the uncertainty relation, the
covariance matrix which represents a physical state should satisfy
\begin{align}
        V+\frac{i}{2}\,\Omega\ge0,
\end{align}
where the inequality of the matrix stands for positive definiteness of
the matrix~\cite{SMD}.  With this property of a physical density
matrix, the PPT criterion is equivalent to the following statement: If
the state is separable, the covariance matrix $\widetilde{V}$ with the
partially transposed density matrix satisfies
\begin{equation}
        \widetilde{V}+\frac{i}{2}\,\Omega\ge0.
\end{equation}
The covariance matrix $\widetilde V$ is easily calculated by inverting
the sign of the momentum $p_i\to-p_i$, which corresponds to partially
transposition of a mode~\cite{Simon}.  By diagonalization of
$\widetilde V$ using a symplectic matrix $S_{d}$,
\begin{align}
        \widetilde{V}+\frac{i}{2}\,\Omega
        =S_{d}^T\left(\bigoplus_i\begin{pmatrix}
                \kappa_i&i/2\\
                -i/2&\kappa_i
        \end{pmatrix}\right)S_{d},
\end{align}
where $\{\kappa_i\}$ are symplectic eigenvalues of $\widetilde{V}$. If
all of the symplectic eigenvalues are greater than $1/2$,
$\widetilde{V}+(i/2)\Omega$ is positive definite.  To quantify
entanglement, the negativity is defined by
\begin{align}
        N=\frac{1}{2}\,\text{max}\left[\left(\prod_{\kappa_i<1/2}\dfrac{1}{2\kappa_i}\right)-1,0\right]
\end{align}
and the logarithmic negativity $L_N:=\log(2N+1)$.  If $N>0$ or
$L_N>0$, the bipartite state is entangled.  Logarithmic negativity is
entanglement monotone (does not increase under the LOCC) and additive,
thus logarithmic negativity can be used as a entanglement
measure~\cite{Vidal,Plenio}.

\section{power spectrum in  low frequency region}
\label{sec:ap-B}
For the trans-sonic case, we expand the Bogoliubov coefficients as a
power series of $\omega$ by comparing the same order terms in the both
sides of the matching formula. For simplicity, we neglect the
$uv$-mixing.  The linear combination of the $u_1$-mode, $u_2$-mode,
and $\bar{u}$-mode is chosen so that it decays exponentially as
$x\rightarrow-\infty$.  We consider wave functions
$A \exp{i k^-_\text{decay}\,x}$ for $x<0$ and
$\exp{i k^+_{u_1}\,x}+B \exp{i k^+_{u_2}\,x}+C \exp{i k^+_{\bar{u}}\,x}$ for
$x>0$, and match them at $x=0$. The matching formula corresponding to
Eq.~\eqref{eq:matching} is written as
\begin{align}
    A=1+B+C,\quad A\, k^-_\text{decay}= k^+_{u_1}+B\, k^+_{u_2}+C\, k^+_{\bar{u}},\quad A (k^-_\text{decay})^2=(k^+_{u_1})^2+B (k^+_{u_2})^2+C (k^+_{\bar{u}})^2.
    \label{eq:matching2}
\end{align}
We expand $A,B,C$ in the power of $\omega$ as
\begin{align}
  A=A^{(0)}+A^{(1)}\omega+\cdots, \quad B=B^{(0)}+B^{(1)}\omega +\cdots, \quad C=C^{(0)}+C^{(1)}\omega+\cdots.
    \label{eq:series}
\end{align}
We substitute Eq.~\eqref{eq:series} into Eq.~\eqref{eq:matching2}, and
equate terms with the same powers of $\omega$.  Wave numbers
$k^-_\text{decay}, k^+_{u_1}, k^+_{u_2}, k^+_{\bar{u}}$ are determined as
solutions of the dispersion relation Eq.~\eqref{eq:dispersion} as
\begin{align*}
    k^-_\text{decay}&=i k_0 \sqrt{V_-^2-1}+\left(\frac{d\Omega}{d k}\right)^{-1}\omega+\cdots,\quad
    k^+_{u_2}=k_0 \sqrt{1-V_+^2}+\left(\frac{d\Omega}{d k}\right)^{-1}\omega+\cdots,\\
     k^+_{\bar{u}}&=k_0 \sqrt{1-V_+^2}+\left(\frac{d\Omega}{d k}\right)^{-1}\omega\cdots,\quad\quad
     k^+_{u_1}=\frac{\omega}{1-V_+}.
\end{align*}
By substituting these expressions into Eq.~\eqref{eq:matching2}, we
obtain coefficients of the wave function in the lowest order of
$\omega$ as
\begin{align}
  A^{(0)}&=1-\frac{V_-^2-1}{V_-^2-V_+^2},\\
  B^{(0)}&=-\frac{1}{2}\left(\frac{V_-^2-1}{V_-^2-V_+^2}-i \sqrt{\frac{V_-^2-1}{1-V_+^2}}\left(1+\frac{V_-^2-1}{V_-^2-V_+^2}\right)\right),\\
  C^{(0)}&=-\frac{1}{2}\left(\frac{V_-^2-1}{V_-^2-V_+^2}+i \sqrt{\frac{V_-^2-1}{1-V_+^2}}\left(1+\frac{V_-^2-1}{V_-^2-V_+^2}\right)\right).
\end{align}
In the zeroth order of $\omega$,
$N^+_{u_2}=N^+_{\bar{u}},
~|\alpha_{u_1u_2}/\alpha_{u_1\bar{u}}|=|\beta_{u_1u_2}/\beta_{u_1\bar{u}}|=|B^{(0)}/C^{(0)}|=1$
holds.


%

\end{document}